\documentclass[5p,twocolumn,final]{elsarticle}
\usepackage{algorithmic}
\usepackage{graphicx}
\usepackage{textcomp}
\usepackage{xcolor}
\usepackage{bm}
\usepackage{amsmath}
\usepackage{amssymb}
\usepackage{amsfonts}
\usepackage{enumitem}

\usepackage[colorlinks]{hyperref}
\makeatletter
\def\ps@pprintTitle{%
\def\@oddhead{\hfill\textit{\@date}} 
}
\makeatother

\begin{document}

\title{A Hybrid Framework for Statistical Feature \\ 
Selection and Image-Based Noise-Defect Detection\tnoteref{t1}}

\author[1]{Alejandro Garnung Menéndez, M.Sc., PhD Candidate}
\address[1]{University of Oviedo, Gijón Polytechnic School of Engineering, Asturias, España}
\ead{uo269564@uniovi.es, garnungalejandro@gmail.com}

\begin{abstract}
In industrial imaging, accurately detecting and distinguishing surface defects from noise is critical and challenging, particularly in complex environments with noisy data. This paper presents a hybrid framework that integrates both statistical feature selection and classification techniques to improve defect detection accuracy while minimizing false positives. The motivation of the system is based on the generation of scalar scores that represent the likelihood that a region of interest (ROI) is classified as a defect or noise. We present around 55 distinguished features that are extracted from industrial images, which are then analyzed using statistical methods such as Fisher separation, chi-squared test, and variance analysis. These techniques identify the most discriminative features, focusing on maximizing the separation between true defects and noise. Fisher’s criterion ensures robust, real-time performance for automated systems.

This statistical framework opens up multiple avenues for application, functioning as a standalone assessment module or as an \textit{a posteriori} enhancement to machine learning classifiers. The framework can be implemented as a black-box module that applies to existing classifiers, providing an adaptable layer of quality control and optimizing predictions by leveraging intuitive feature extraction strategies, emphasizing the rationale behind feature significance and the statistical rigor of feature selection. By integrating these methods with flexible machine learning applications, the proposed framework improves detection accuracy and reduces false positives and misclassifications, especially in complex, noisy environments. 
\end{abstract}


\maketitle

\section{Introduction}
The detection and classification of noise in images is crucial for numerous industrial applications, particularly in automated inspection systems. Image processing has always highlighted the need for effective noise classification methods that can distinguish between noise and actual defects. Many works address the intricate problem of differentiating between noise and actual surface defects in industrial parts, which frequently requires in-depth statistical analysis to uncover subtle differences that are not immediately apparent \cite{MariaInesSilva}.
Moreover, feature selection plays a critical role in enhancing machine learning (ML) performance by reducing false positives. This article focuses on statistical methods for feature selection in image-based defect detection and the usage of these enhanced features that help build an effective discriminative method for defect and noise classification. These techniques aim to identify the most effective features that can distinguish between true positives or true defects (TP) and false positives or false defects (FP) in detection tasks. Specifically, the aim is to develop a dichotomous image classification methodology based on specific scalar features extracted from candidate images. The proposed hybrid framework acts as a black-box model that integrates multiple analytical methods which can be assembled on top of existing models to classify ROIs as noise or defects. For example, it enables the evaluation of the outputs of other models using simple threshold-based methods or by training a model such as a random forest (RF).

On the one hand, this study presents a comprehensive methodology for feature selection aimed at optimizing the separation between TP and FP in a defect detection system, specialized in the morphology of highly noisy images degraded by random and impulse noise of defects captured with laser sensors. The analysis begins with a set of approximately 55 extracted features, seeking to identify those that best differentiate between the two classes while ensuring long-term reliability. The primary objective is to retain the features where the distributional means of TP and FP are maximally separated, to allow efficient and straightforward threshold-based classification. Using statistical methods, our aim is to develop a robust feature selection process that can be applied to various supervised, unsupervised, or hand-crafted classification methods, enhancing their precision and generalizability across different types of defects and environments. This integration, based on statistical intuition, enhances the system's adaptability to complex real-world scenarios, ensuring reliability across diverse applications.

On the other hand, we present a novel approach that leverages this wide range of analytical metrics, combining them into a simple yet ad hoc weighted scoring system. One could further enhance and generalize this system using machine learning models to optimize decision making. This capitalizes on inherent advantages: robustness to noise, ability to handle complex datasets, flexibility in managing various types of input characteristics, and feature selection based on its relative importance. 

In this context, our primary concern is that both noise and defects occupy a very small portion of the dynamic range, resulting in inherently low contrast, which makes classification challenging. However, we can capitalize on the fact that our type of noise is often accompanied by point losses, typically represented by values below the three-bit digital threshold. We aim to characterize and differentiate the structure and arrangement of these losses, as they seem to exhibit distinct patterns compared to the defects analyzed in this study.

\section{Related work}
The objective of this article is to contribute to the literature by providing a repository of useful methods for the detection and classification of noise under particularly unfavorable scanning conditions where defect detection (such as cracks, dents, scratches, chips, etc.) is required. In these scenarios, the geometric structure of the noise and the defects themselves is often indistinguishable with the naked eye.

In \cite{ZhifeiZhang} the author reviews the evolution of image noise reduction techniques, focusing on detection, measurement, and removal, with an emphasis on noise measurement and denoising methods. The work explores measurement approaches such as filter-based, block-based, and wavelet-transform methods, along with other significant advancements. Noise removal techniques are discussed across four historical phases, from the 1970s to the early 2010s, transitioning through spatial and transform domains, local and non-local statistical methods, and recent innovations like total variation minimization, sparse coding, and deep learning. The article highlights a shift from spatial to transform domains, from local to non-local approaches, and from thresholding to optimization. Despite the evolution of techniques, the primary goal remains unchanged: smoothing images while maintaining their essential details.

The authors in \cite{YiqiuDong} introduce a new image statistic designed to detect random-valued impulse noise in images. This statistic enables the identification of most noisy pixels in corrupted images, and when combined with an edge-preserving regularization, it forms a powerful two-stage denoising method that effectively handles high noise levels. The effectiveness of the method has been shown to reduce false detection of noisy pixels while maintaining high image quality, but remains restricted to a denoising application.

Recently, in \cite{RusulMudhafar} a method for detecting and classifying various types of image noise, such as Gaussian, log-normal, Rayleigh, salt and pepper and speckle \cite{Mudhafar2024} was proposed. The method uses a two-stage approach: first, a Convolutional Neural Network (CNN) detects the noise, and then machine learning classifiers, combined with a deep wavelet model, classify the noise.

Moreover, \cite{Tang2015} proposes a framework for estimating the noise level in natural images using statistical analysis, focusing on high kurtosis and scale invariance in the transform domain, examining the limitations of these statistics, especially for images with highly directional edges or large smooth areas. They combine wavelets, non-directional discrete cosine transform (DCT) and noise-injection rectification to estimate noise variance across a wide range of image content and noise levels. However, the scope of the method may be limited to enhancing denoising.

In \cite{ElvanDUMAN}, it is presented a statistical edge detection method designed to be more robust to noise, which can degrade the performance of segmentation and edge detection. Traditional methods are vulnerable to noise, while statistical tests like the t-test, Wilcoxon test, and rank-order test have been proposed for noisy images. The authors introduce a new approach that combines the rank-order test with k-means clustering to enhance its efficiency. 

Similarly, \cite{Kim2005} presents algorithms for distinguishing image features from noise in video processing applications, especially in environments with time-varying noise, such as digital TV. The difficulty lies in separating noise from the signal, as they are hard to distinguish. The proposed algorithms use statistical hypothesis testing to measure the degree of noise relative to the image feature, based on the random nature of noise and the spatial correlation of image features. The framework evaluates the randomness of noise through a statistical measure and does not rely on assumptions like homogeneous areas or prior knowledge of the noise strength. However, not relying on prior information, no matter how minimal, can be detrimental rather than beneficial for applications such as defect detection and robustness against types of noise.

To the best of our knowledge, there is limited literature on the general detection of noise and subsequent decision-making based on the type of detected noise in the sense in which we think of approaching it, and most studies primarily focus on Gaussian or impulsive random noise. Our aim is to distinguish more structured forms of noise that are not solely caused by electronic noise or white noise interference; in fact, those would be issues that could be relatively easily mitigated with simple diffusion filters and edge preservation techniques. Instead, we propose addressing the detection of data loss-related noise and more coarse-grained noise types that could easily be misinterpreted as actual defects.

Additionally, it is important to acknowledge that, due to the environment, time, or resource constraints, upgrading sensing technology or instrumentation is not always feasible. This limitation necessitates the development of robust methodologies capable of handling data that are far from ideal. This work serves as an example of striving to overcome such challenges by designing effective solutions tailored to suboptimal conditions.

To this end, we address these problems from a different perspective, grounding out work in statistical and applied mathematical methods, aiming to provide a framework for developing efficient algorithms for noise classification and detection.

\section{Methodology}
The primary goal of the methodology is to provide robust and reliable features to an image classification system for dichotomous tasks. This involves generating a diverse monochrome or vector-valued dataset from the ROIs of the raw images, heuristically establishing mathematical models for noise and defects and extracting and analyzing features to distinguish the key ones that best separate noise and defects. Additionally, the methodology includes rigorous statistical feature selection posed to an ulterior application of aggregation methods, such as weighted score approaches or RF models, all aimed at classification accuracy and generalization in complex industrial scenarios.

\subsection{Dataset generation}
\begin{figure}[h!]
    \centering
    \includegraphics[width=\linewidth]{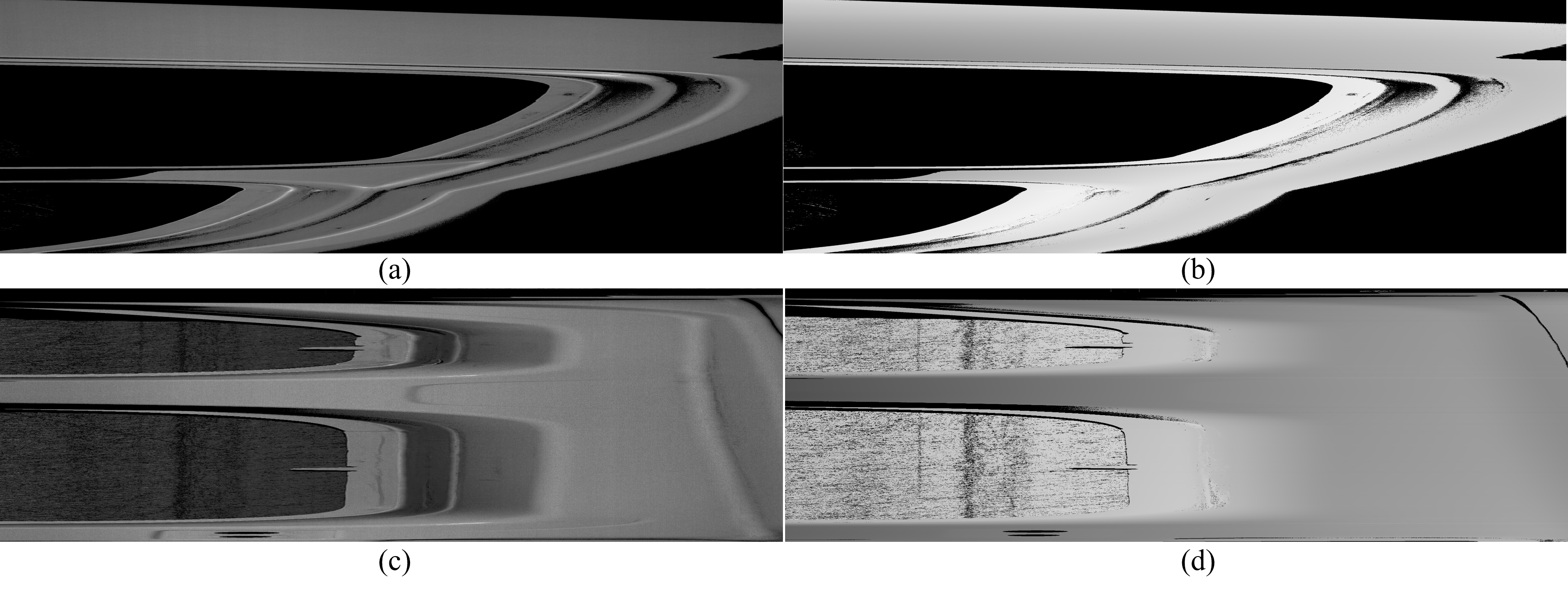}
    \caption{a, c) Luminance images of scanned parts. b, d) Corresponding distance images. These images are highly noisy, making it impossible to visually discern from this data whether they are defective or not.}
    \label{fig:Dataset1}
\end{figure}

The framework developed is versatile and can be applied to a broad range of image types, including color images, monochrome images, luminosity maps, distance images, and depth maps. This flexibility allows it to be used in various fields and in real-world applications. Specifically, the dataset used for testing is tailored for industrial applications and consists of images from raw three-channel data (luminance, normal or distance maps) (see Fig.~\ref{fig:Dataset1}), with a focus on variations in distance and lighting captured by laser triangulation profilometers. These images include defective and nondefective industrial images, with varying levels of noise, ranging from low to high density. This diversity in noise levels increases the dimensionality of the input space of our model and improves its ability to generalize between different conditions that can arise in practical use cases.

We consider a wide variety of types of defects, such as fractures, dents, and tears in steel components, within a dynamic industrial environment, focusing on key aspects such as luminosity, intensity distribution, and statistical pixel metrics, which are crucial for effective noise and defect detection. Advanced ML techniques were used to accurately segment the candidate ROIs for testing the proposed features, allowing the precise extraction of areas that potentially contain noise or defects. The varied content including complex lighting conditions, different types of noise and real defects provides a comprehensive foundation for training robust models capable of handling a wide range of real-world challenges in image analysis.

\begin{figure}[h!]
    \centering
    \includegraphics[width=\linewidth]{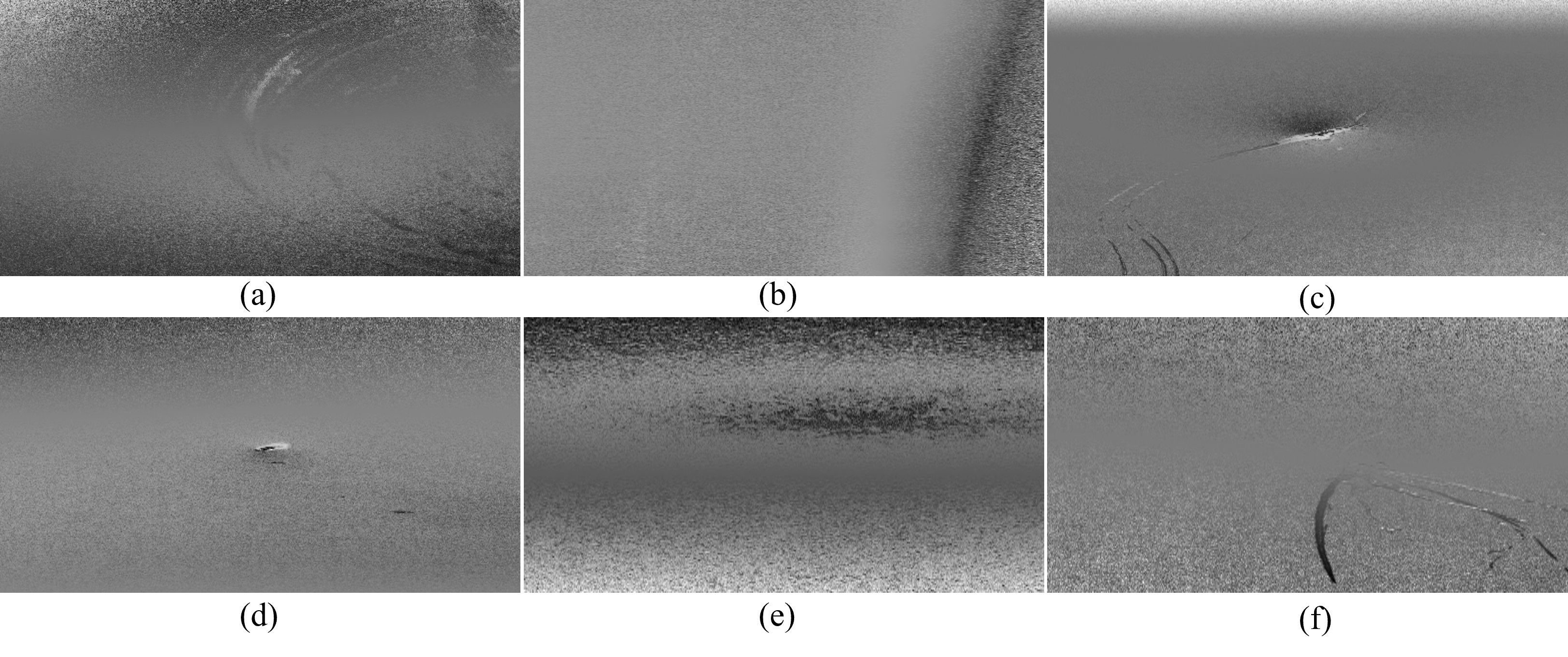}
    \caption{a-f) All images are defective but were classified as noisy by a classification model.}
    \label{fig:tps}
\end{figure}

\begin{figure}[h!]
    \centering
    \includegraphics[width=\linewidth]{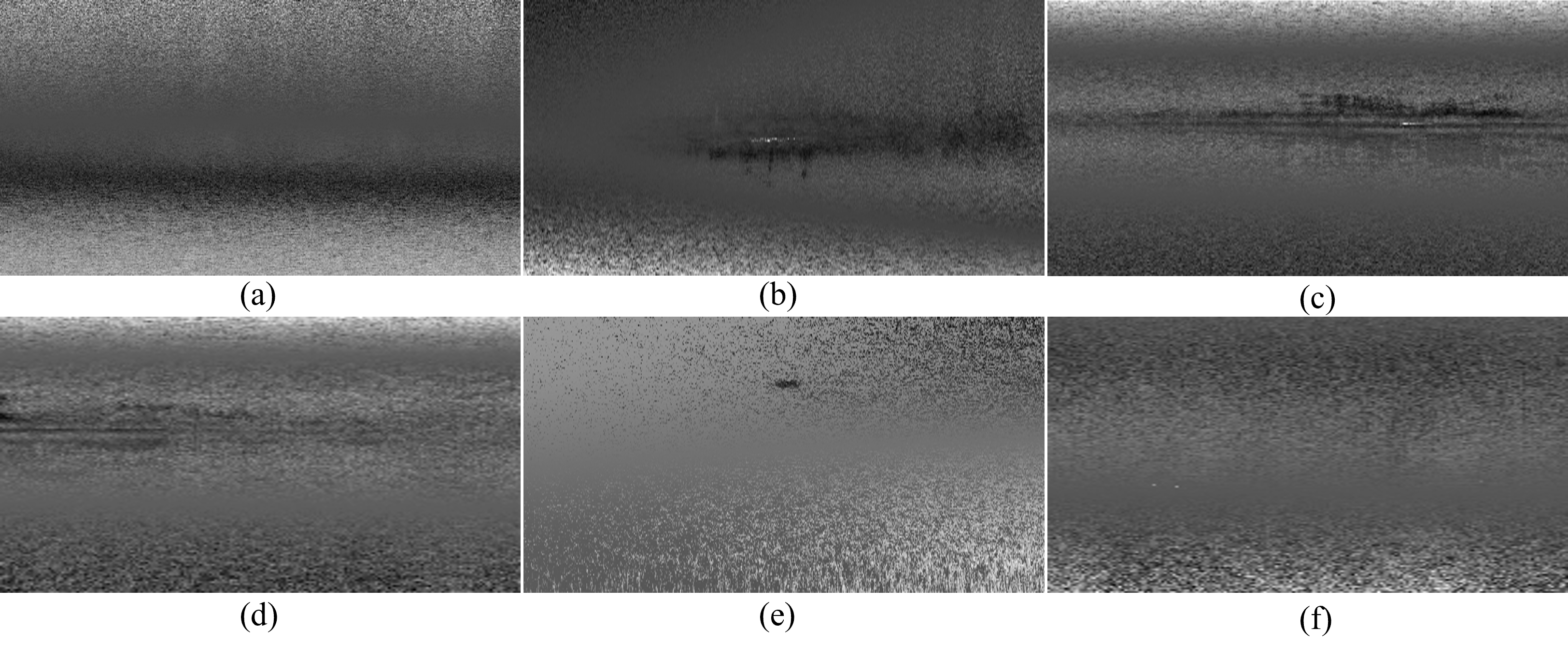}
    \caption{a-f) All images are noisy but were classified as defective by a classification model.}
    \label{fig:fps}
\end{figure}

In Fig.~\ref{fig:tps} and Fig.~\ref{fig:fps}, the complexity of the scenarios for classifying between noise and defects is evident. It is necessary to develop a robust framework to address the high variability of the data from a statistical perspective.

\subsection{Feature Extractors}
In this first stage, we focus on identifying basic and advanced features that can effectively separate different types of defects from noise. We employ a variety of discriminative methods and metrics for feature extraction, leveraging intuitive processing strategies, and emphasizing the rationale behind its significance. The features are organized into several categories on the basis of their mathematical principles.

We analyze spatial relationships using techniques that assess the structural integrity and arrangement of pixels to identify significant patterns amid noise. We also examine distribution measures through statistical testing and histogram comparison metrics, which help identify anomalies in pixel distributions that may indicate noise. Additionally, we utilize texture extraction methods to differentiate valid content from random noise by examining the consistency of pixel arrangements and to quantify pixel noise variability. Further, we incorporate directly threshold techniques to decision, distance metrics between detected features, probabilistic modeling, and clustering analysis to help describing the underlying distributions of pixel values and identifying meaningful groupings.

\subsection{Statistical feature selection}
To ensure the relevance of the selected candidates, in this second stage, we prepare a framework to identify the most distinctive characteristics that explain the nature of the data set. In this second stage, a single workflow may be executed to retain the best extracted features, separating the classes in the most effective way, assessing their statistical distribution to ensure long-term reliability and filtering the best ones based on various criteria.

The main basic idea is to retain the features whose means for the FP and TP cases are the most separated, facilitating defect separation through threshold techniques. Additionally, necessary assumptions are made to enhance the stability and efficiency of the block, aiming to maximize inter-class variability and minimize intra-class variability to improve discernment capabilities, supporting real-time threshold-based decision-making with improved generalization across diverse defect types and operational scenarios.

Two consecutive approaches are used for feature selection. The first targets features that approximate a uniform distribution by maximizing the empirical mean difference between classes while maintaining low variance. The second focuses on features resembling a normal distribution, applying stringent conditions such as normality checks, the Kolmogorov-Smirnov (K-S) test, and the Student's t-test. This step ensures that selected features align closely with a normal distribution and satisfy established criteria, maximizing inter-class variability and minimizing intra-class variability based on Fisher's criterion. Finally, the Bhattacharyya distance is calculated to evaluate classification quality, retaining only features that surpass a specified threshold, which helps define the boundaries of each class.

\subsection{Aggregating methods}
In a final stage, one aims to test the framework on-field. We primarily explore an aggregating systematic and simple yet effective method that combines the selected features based on a scoring approach. A brief test of the framework and most of the proposed features is performed, including evaluating it on a subset of the image data and validating its performance on unseen data, ensuring the well generalization beyond the training set. We also mention the possibility of using learning models, like RF, which enables the usage of optimal weights for each analytical metric, enhancing overall performance, and mitigating overfitting.

\section{Feature extractors}
In our framework, we extract scalar features from images that capture essential and intrinsic statistical and structural properties. Using our dataset, we perform a rigorous and extended analysis, identifying advanced features that can effectively separate different defect types from noise is inherently challenging. Therefore, we outline discriminative methods and metrics for feature extraction employing intuitive processing strategies.

In this section, we explore the discriminative methods employed in our model to effectively identify and mitigate noise within the image data. Each method is discussed individually, along with its intuitive justification. We can categorize the proposed features into distinct groups based on their characteristic nature and mathematical intuition.
\begin{itemize}[label={-}]
    \item {Spatial relationships:} Methods like connected component analysis and centroid proximity filtering are used to evaluate the structural integrity and spatial arrangement of pixels. These techniques help discern meaningful patterns in the image by analyzing how the components of the pixels relate to each other, allowing the identification of significant structures among the noise.
    
    \item {Measures of distributions:} This category includes features such as statistical testing for the uniformity of the pixel value distribution and histogram comparison metrics. These techniques assess the statistical characteristics of the pixel value distributions, helping to identify anomalies that may indicate noise. By understanding the distribution of pixel values, we can discern between typical and atypical data points in an image.
    
    \item {Texture extraction:} We utilize texture homogeneity assessment and spatial clustering analysis of the pixel value distribution. These methods focus on identifying patterns and textures within the image. They help differentiate meaningful image content from random noise by analyzing the consistency and coherence of pixel arrangements, thereby enhancing the model's ability to recognize valid features.
    
    \item {Variance and statistical analysis:} This group involves techniques such as Variance Analysis through Interquartile Range (IQR) and Gaussian Mixture Model (GMM) fitting. These methods quantify the variability and distribution characteristics of the pixel values, enabling us to detect noise and distinguish consistent patterns from random fluctuations.
    
    \item {Threshold techniques:} This category includes global and adaptive umbralization, which apply decision thresholds to filter significant features from noise, enhancing the overall detection capability of the model by ensuring that only relevant data points are considered.
    
    \item {Distance metrics:} Techniques that involve distance metrics for proximity analysis are utilized to measure spatial relationships between detected features. By analyzing distances, we can determine the relevance of detected points, ensuring that only closely related candidates are selected for further consideration.
    
    \item {Probabilistic modeling:} This group includes techniques such as fitting probabilistic models for histogram analysis, which leverage statistical models to describe the underlying distributions of pixel values, providing a robust framework for understanding image structure and identifying noise characteristics.
    
    \item {Clustering analysis:} In this context, clustering techniques are used to analyze spatial arrangements of pixel values, helping to identify groupings that may represent meaningful structures within the image while filtering out random noise.
\end{itemize}

\subsection{Absolute Pixel Count Evaluation}
Let \( S \) be the set of all pixels in the scanned image such that \( |S| = N \), where \( |S| \) denotes the cardinality of the set \( S \).

Define the subsets \( S_R \) and \( S_D \): \( S_R \subset S \) is the subset of pixels classified as noise and \( S_D \subset S \) is the subset of pixels classified as defects.

Let \( L \) represent the set of lost pixels, defined by \( L = S \setminus \left( S_R \cup S_D \right) \) and let \( a_0 = |L| \) denote the absolute count of lost pixels.

To distinguish between noise and defects, we introduce a probabilistic framework. Define \( P(N) \) as the probability that a pixel is classified as noise, \( P(D) \) as the probability that a pixel is classified as a defect, and \( \lambda \) as a threshold parameter that determines the critical pixel loss ratio. We can express the relationships between these probabilities as:
\[
    P(N \!\mid\! a_0) = \frac{P(a_0 \!\mid\! N) P(N)}{P(a_0)},
\]
\[
    P(D \!\mid\! a_0) = \frac{P(a_0 \!\mid\! D) P(D)}{P(a_0)}.
\]

Let the previous equations represent the conditional probabilities of observing \( a_0 \) given that the pixels are classified as noise or defects, respectively. Heuristically, we assume that:
\[
P(a_0 \!\mid\! D) > P(a_0 \!\mid\! N).
\]
This indicates that the expected number of lost pixels is greater when defects are present than when noise is present. This aligns well with empirical measurements obtained from imaging applications.

To identity the class, we define a decision criterion based on the absolute pixel count:
\[
    C(a_0) = 
    \begin{cases} 
        D & \text{if } a_0 \geq \lambda \\
        N & \text{if } a_0 < \lambda,
    \end{cases}
\]
where \( C(a_0) \) is the classification function that returns either \( D \) (where \( D = 1 \) indicates a defect) or \( N \) (where \( N = 0 \) indicates noise) based on comparison of \( a_0 \) to the threshold \( \lambda \), which can be based on the distributions of pixel loss counts under noise and defects.

\subsection{Expected Absolute Pixel Count Evaluation}
The expected absolute pixel count under the assumption of noise versus defects can be modeled as:
\[
    E[a_0 \!\mid\! D] = \mu + \sigma_D^2, 
\]
\[
    E[a_0 \!\mid\! N] = \mu + \sigma_N^2.
\]
where \( \mu_D \) and \( \mu_N \) represent the mean pixel loss counts when there are defects and noise, respectively. These values indicate the average number of pixels expected to be lost under each condition. Similarly, \( \sigma_D^2 \) and \( \sigma_N^2 \) represent the variances of the pixel loss counts under defects and noise, respectively. Although variance provides insight into the spread and consistency of pixel loss counts around the mean, it does not directly influence the expected value.

In classical probability theory, the expected value refers to a weighted average of all possible outcomes, calculated based on their probabilities. In our context, we focus solely on the mean pixel counts associated with specific conditions (defects or noise).

To create a decision metric that uses these expected values, we can employ a statistical hypothesis testing framework, so we can define another decision metric \( C(a_0) \).

Using likelihoods based on expected values and a threshold, we can derive:
\[
    C(a_0) = 
    \begin{cases} 
        D & \text{if } \; \dfrac{L(D \mid a_0)}{L(N \mid a_0)} > \lambda \\[8pt]
        N & \text{if } \; \dfrac{L(D \mid a_0)}{L(N \mid a_0)} \leq \lambda, 
    \end{cases}
\]
where \( L(D \!\mid\! a_0) \) is the likelihood of observing \( a_0 \) since it is a defect, and \( L(N \!\mid\! a_0) \) is the likelihood of observing \( a_0 \) given that it is noise.

Assuming a normal distribution, the likelihoods can be expressed in terms of the expected values:
\[
    L(D \mid a_0) \propto \exp\left(-\frac{(a_0 - E[a_0 \!\mid\! D])^2}{2\sigma_D^2}\right),
\]
\[
    L(N \mid a_0) \propto \exp\left(-\frac{(a_0 - E[a_0 \!\mid\! N])^2}{2\sigma_N^2}\right).
\]

Substituting the expected values, we have:
\[
    L(D \mid a_0) \propto \exp\left(-\frac{(a_0 - (\mu_D + \sigma_D^2))^2}{2\sigma_D^2}\right), 
\]
\[
    L(N \mid a_0) \propto \exp\left(-\frac{(a_0 - (\mu_N + \sigma_N^2))^2}{2\sigma_N^2}\right).
\]

Putting it all together, the decision metric can be reformulated as:
\[
    C(a_0) = 
    \begin{cases} 
        D & \text{if } \; \exp\left(-\frac{(a_0 - (\mu_D + \sigma_D^2))^2}{2\sigma_D^2}\right) > \lambda \\[8pt]
        N & \text{if } \; \exp\left(-\frac{(a_0 - (\mu_N + \sigma_N^2))^2}{2\sigma_N^2}\right) \leq \lambda.
    \end{cases}
\]

This integration of expected values into the final decision metric allows for a rigorous and statistically grounded approach to distinguishing between noise and defects based on the absolute pixel count of lost pixels. In addition, the use of likelihood ratios and the consideration of statistical properties of the data enhance the robustness of the decision-making process.

\subsection{Statistical Testing for Pixel Value Distribution Uniformity}
The goal of this method is to analyze the distribution of black pixels in ROIs to determine whether they follow a uniform distribution, which would indicate the presence of impulsive noise, or whether they are clustered, suggesting the existence of defects. The following hypothesis-based strategy is employed:
\begin{itemize}[label={-}] 
    \item Null Hypothesis (\(H_0\)): The distribution of black pixels in the ROI follows a uniform distribution, implying that black pixels are the result of impulsive noise.
    \item Alternative Hypothesis (\(H_1\)): The distribution of black pixels is more dispersed in the noisy ROI than in the defect ROI, indicating that the black pixels are more spatially clustered.
\end{itemize}

The selected ROI is divided into \( N \) patches \( P_i \) of size \( s \times s \). Only patches containing at least \( \varepsilon \in \mathbb{N}^* \) black pixels are considered, so the total number of black pixels in ROI is given by:
\[
T = \sum_{i=1}^{N} B_i \quad \text{for } B_i \geq \varepsilon,
\]
where \( B_i \) is the number of black pixels in patch \( P_i \).

The expected number of black pixels in each patch, under the assumption of a uniform distribution, is calculated as \( E = T \mathbin{/} N \), and the observed counts are denoted as \( O_i \) for each patch \( P_i \).

The chi-square test is applied to evaluate the discrepancy between the observed and expected values. It is calculated as:
\[
\chi^2 = \sum_{i=1}^{N} \frac{(O_i - E)^2}{E}.
\]

The associated \( p \)-value is obtained from the chi-square distribution with \( N-1 \) degrees of freedom. A \(p\)-value \textless{} \(\alpha\) (where \( \alpha \) is the significance level, commonly set at 0.05) leads to the rejection of \( H_0 \), suggesting that the black pixels do not follow a uniform distribution.

The analysis is complemented by the Monte Carlo and Fisher tests, calculating their corresponding \( p \)-values to validate the results obtained. The Monte Carlo test estimates the distribution of a test statistic under the null hypothesis by simulating a large number of random samples. Mathematically, this process can be defined as follows:
\begin{itemize}[label={-}]
    \item To conduct the Monte Carlo test, we first generate \( N \) random samples from a Poisson distribution, which models the number of events occurring in a fixed interval of space. In this context, the “events” refer to the number of black pixels within a patch of the image. The Poisson distribution is characterized by its probability mass function, defined as:
    \[
    P(X = k) = \frac{\lambda^k e^{-\lambda}}{k!},
    \]
    where \( \lambda \) represents the average number of occurrences (the expected number of black pixels per patch), denoted as \( E \). Here, \( X \) is the random variable that indicates the number of black pixels in a given patch, and \( k \) is the specific count of occurrences (i.e., the number of black pixels).
    \item For each sample \( X_i \), we compute the test statistic \( \chi_i^2 \) using the formula:
    \[
    \chi_i^2 = \sum_{j=1}^{k} \frac{(X_{ij} - E)^2}{E}.
    \]
    \item The observed test statistic \( \chi_{\text{o}}^2 \) is calculated as:
    \[
    \chi_{\text{o}}^2 = \sum_{j=1}^{k} \frac{(O_j - E)^2}{2},
    \]
    where \( O_j \) denotes the observed count of black pixels in each patch.
    \item The \( p \)-value is finally computed as:
    \[
    p\text{-value} = \frac{1}{N} \sum_{i=1}^{N} \mathbf{1}_{(X_i^2 \geq X_{\text{o}}^2)},
    \]
    where \(\mathbf{1}_{(.)}\) is the indicator function, that equals 1 if the condition is true and 0 otherwise.
    
    \item Moreover, the Fisher test is employed to assess the statistical significance of associations between two categorical variables, typically structured in a contingency table. This can be expressed mathematically as follows:
    Let \( a \) be the count of patches with black pixels, \( b \) be the count of patches without black pixels, \( c \) be the count of patches in the defect category, and \( d \) be the count of patches without defects. The contingency table is arranged as follows:
    \[
    \begin{array}{|c|c|c|}
    \hline
    \text{Black Pixels} & \text{Defects} & \text{Count} \\
    \hline
    \text{Yes} & \text{Yes} & a \\
    \hline
    \text{Yes} & \text{No} & b \\
    \hline
    \text{No} & \text{Yes} & c \\
    \hline
    \text{No} & \text{No} & d \\
    \hline
    \end{array}
    \]

    \item The probability of obtaining the observed counts under the null hypothesis (which posits independence between the two categories) is computed using the hypergeometric distribution:
    \[
    p = \frac{\binom{a+b}{a} \cdot \binom{c+d}{c}}{\binom{N}{n}},
    \]
    where \( N = a + b + c + d \) is the total number of patches and \( n = a + b \) is the total number of patches with black pixels.

    \item A \( p \)-value is derived based on the observed counts. If \( p < \alpha \) (commonly set at 0.05), it indicates a significant association between the presence of black pixels and the occurrence of defects.
\end{itemize}
An effective configuration for patch distribution was to utilize 4 or 8 patches per row and column, particularly with images at a downsampling size of 512x512 (this can help mitigate noise and enhance the impulsivity of defect detection). Although the chi-square test is a useful tool, it requires an extremely low \( p \)-value (e.g., \(p\!<\!0.01\)) to produce significant results, indicating a very high confidence level. 

\subsection{Variance analysis via IQR measurement}
The IQR is a robust measure of statistical dispersion, especially useful in image processing to detect deviations in the intensity distributions of pixels. In the context of image processing, the IQR can serve as an indicator of anomalous pixel values, helping to differentiate between noise and defects. 

Given an image of size \( M \times N \), we define the set of all pixel values as the 1-vector:
\[
S(u) = \{ u(i,j) \, | \, i = 1, \ldots, M; \, j = 1, \ldots, N \}.
\]

Next, we sort the pixel values in nondecreasing order, resulting in a set \( S_o(u) \), where 
\[
S_o(u) = \{ S_{o_1} \leq S_{o_2} \leq \ldots \leq S_{o_{M \cdot N}} \},
\]
where \( S_{o_k} \) represents the \( k \)-th pixel value in the ordered list, such that \( S_{o_1} \) is the smallest and \( S_{o_{M \cdot N}} \) is the largest pixel intensity.

Let \( n = M \cdot N \) denote the total number of pixels; we mathematically define the IQR \( I \) as follows:
\[
IQR = S_{o\left(\frac{3}{4}(n+1)\right)} - S_{o\left(\frac{1}{4}(n+1)\right)} = Q_3 - Q_1,
\]
where \( Q_1 \) is the first quartile, representing the 25th percentile of the sorted pixel intensity distribution, and \( Q_3 \) is the third quartile, representing the 75th percentile. 

Outliers, which are candidate noise or defect pixels, are those values that fall outside a specified range defined by the IQR. Specifically, a pixel value \( u(x, y) \) at spatial coordinates \( (x, y) \) in the image grid domain \( \Omega \) is considered an outlier if it satisfies either of the following conditions:
\[
u(x, y) < Q_1 - a \cdot 1.5 \cdot IQR \quad \lor \quad u(x, y) > Q_3 + b \cdot 1.5 \cdot IQR,
\]
where \( a, b \in \mathbb{R} \) are scaling factors that can be adjusted to control the sensitivity of the outlier detection process. In standard filtering techniques, \( a = b = 1 \) is commonly used, but these parameters can be tuned to adapt to specific noise characteristics or types of defects in the image.

In image processing tasks where distinguishing noise from structural defects is critical, relying solely on the IQR may not be sufficient. To enhance this method, we introduce an adaptive heuristic using local variance as an additional feature to refine noise detection.

\begin{figure}[h!]
    \centering
    \includegraphics[width=\linewidth]{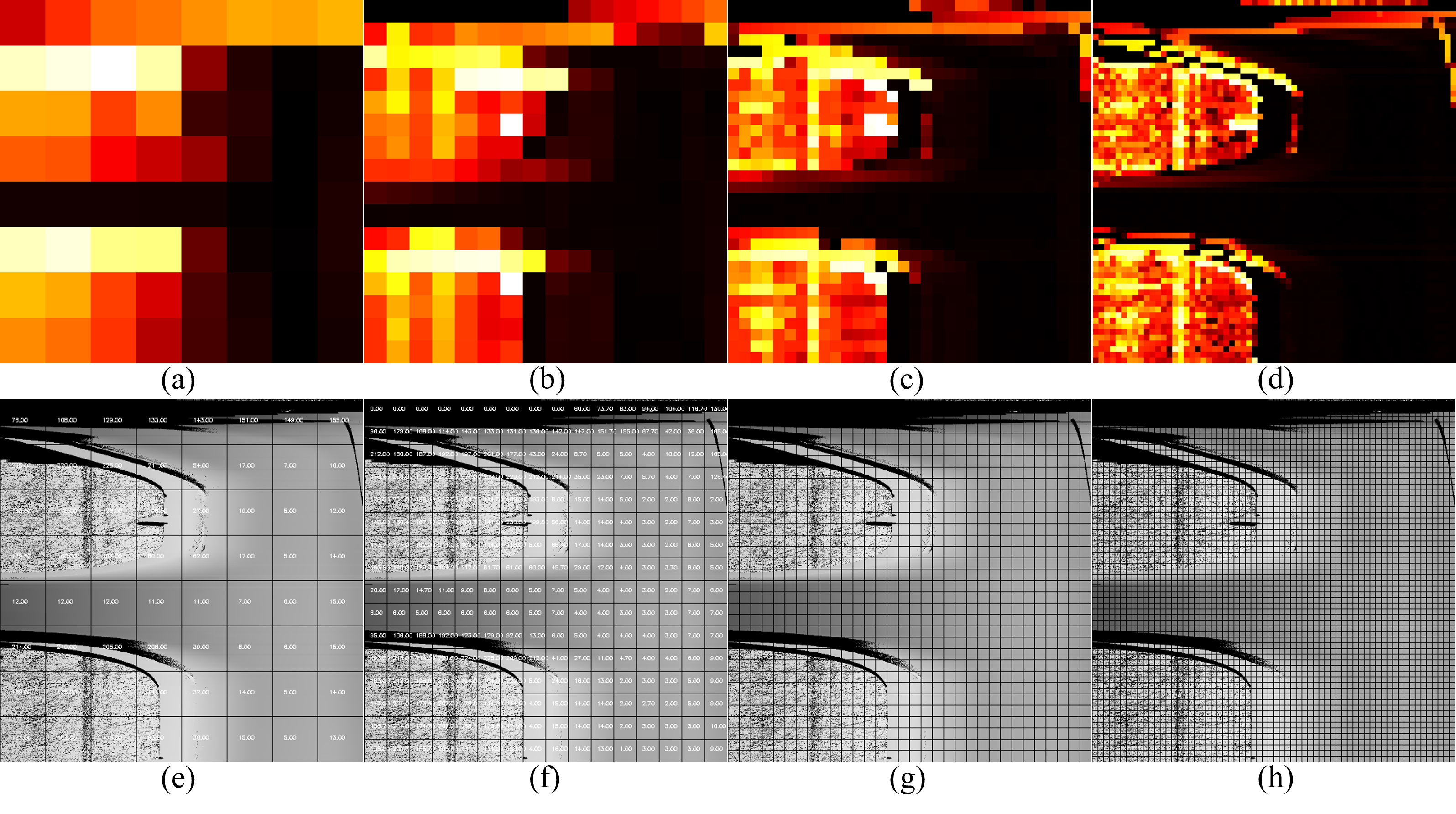}
    \caption{Coarse-to-fine patch-based analysis of IQR neighborhoods in candidate. a-d) Heatmaps highlighting most probable noisy regions. e-h) Metric value computation per patch.}
    \label{fig:patches}
\end{figure}

Let \( \mathcal{N}(x, y) \) represent the neighborhood of the pixel \( u(x, y) \), and let \( \sigma_{\text{local}}^2(x, y) \) denote the local variance of the values of the pixels in this neighborhood. A pixel is more likely to represent noise if the local variance is small, indicating that the value of the pixel is not part of a contiguous anomaly. Therefore, let \( C_{\text{defective}} \) be the class of defective images and \( C_{\text{noise}} \) be the class of noisy images; we introduce the following condition for noise classification:
\[
u(x, y) \in C_{\text{defective}} \quad \text{if} \quad \sigma_{\text{local}}^2(x, y) < \lambda_{\text{noise}},
\]
where \( \lambda_{\text{noise}} \) is a threshold that can be learned from the dataset or determined empirically (see Fig.~\ref{fig:patches}). In contrast, pixels that are part of defects are more likely to exhibit higher local variance due to the structural discontinuities they represent.
This combination of IQR-based outlier detection and local variance filtering allows for a more refined noise discrimination process. Formally, for each pixel \( u(x, y) \), we define the detection criteria by an outlier/noise function as follows:
\[
O(x, y) = 
\begin{cases} 
1, & \text{if } (u(x, y) \leq Q_1 - c_1 \hspace{0.1cm} \lor \\
& \quad \hspace{0.1cm} u(x, y) \geq Q_3 + c_2) \hspace{0.1cm} \land \\ 
& \quad \hspace{0.1cm} \sigma_{\text{noise}}^2(x, y) > \lambda_{\text{noise}} \\ 
0, & \text{otherwise},
\end{cases}
\]
where \( c_1 = a \cdot 1.5 \cdot IQR \) and \( c_2 = b \cdot 1.5 \cdot IQR \).

The pixel \( u(x, y) \) is defined to belong to the set \( D \) if \( O(x, y) = 1 \), which can be mathematically expressed as:
\[
u(x, y) \in D \quad \text{if} \quad O(x, y) = 1.
\]

Next, the local variance filter is applied to pixels identified as outliers:
\[
L(x, y) = 
\begin{cases} 
1, & \text{if } \sigma_{\text{local}}^2(x, y) < \lambda_{\text{noise}} \\
0, & \text{otherwise}.
\end{cases}.
\]

Finally, the pixel \( u(x, y) \) is classified as noise or defect according to the rule:
\[
C(x, y) = 
\begin{cases} 
N, & \text{if } O(x, y) = 1 \land L(x, y) = 1 \\ 
D, & \text{otherwise}.
\end{cases}
\]

\subsection{Gaussian Mixture Model Fitting for Histogram Analysis}
In this approach, the pixel intensity distribution of an image \( u \) is modeled using a GMM, under the assumption that the pixels are independent and identically distributed (\textit{iid}). The goal is to approximate the histogram of pixel intensities within \( u \) as a mixture or weighted aggregation of multiple Gaussian distributions, which can help distinguish between noise, casual pixel variations, and potential defects in the image.

The GMM is a clustering-related technique that describes multimodal probability densities by combining simpler constituent distributions \cite[p. 77]{Prince}. In the GMM, the data (i.e., the pixel intensities \( u(x,y) \)) are described as a weighted sum of \( K \) normal distributions:
\[
Pr(x \!\mid\! \theta) = \sum_{k=1}^{K} \lambda_k \cdot {\mathcal{N}}_x(\mu_k, \Sigma_k),
\]
where \( \mu_1, \ldots, \mu_K \) and \( \Sigma_1, \ldots, \Sigma_K \) are the means and covariance matrices of normal distributions, \( \lambda_1, \ldots, \lambda_K \) are positive-valued weights that sum to one, and \( \mathcal{N}_x(\mu_k, \Sigma_k) \) is the normal distribution of the exponential family.
\[
{\mathcal{N}_x}(\mu_k, \Sigma_k) = \frac{\exp\left(-\frac{1}{2} (x - \mu_k)^\top \Sigma_k^{-1} (x - \mu_k)\right)}{(2\pi)^{\frac{d}{2}} |\Sigma_k|^{\frac{1}{2}}} ,
\]

We can think of the grayscale image as a point in \(\mathbb{R}^{M \cdot N} = \mathbb{R}^n\), where \(n\) is the total number of pixels it contains; each pixel intensity is stored in a different dimension. We first vectorize the image by concatenating all its elements into an \(n\)-dimensional vector. In order to learn the parameters \(\theta = \{ (\mu_k, \Sigma_k, \lambda_k) \}_{k=1}^{K}\) from the training data \(\{x_i\}_{i=1}^n\), we could apply the straightforward maximum likelihood approach. However, in this case, the extremality principle for maximizing it is not solvable in closed form, hence requiring a nonlinear optimization.

Alternatively, we could express the observed density as a marginalization and use the well-known Expectation-Maximization (EM) algorithm to learn the parameters. We choose to use the EM approach because of its simplicity compared to other nonlinear methods. In this way, we fit the GMM to the histogram of the pixel intensities of each class dataset and iteratively optimize the model parameters \(\theta\).

In the Expectation step (E-Step), we compute the posterior probability (responsibility) that each pixel intensity \( x_i \) belongs to each Gaussian component \( k \). Specifically, we calculate the responsibility that the intensity of the pixel \( x_i \) is generated by the normal distribution of the \( k \)-th constituent, which is given by:
\[
P(X|C_k) = \gamma_{m,k} = \frac{w_m \mathcal{N}(p_k \!\mid\! \mu_m, \sigma_m^2)}{\sum_{m'=1}^K w_{m'} \mathcal{N}(p_k \!\mid\! \mu_{m'}, \sigma_{m'}^2)}    .
\]

In the maximization step (M-Step), we make use of the Lagrange multipliers, maximizing the bound with respect to the model parameters \( \theta \) and updating them based on the calculated responsibilities. The update for \( \mu_m \) is:
\[
\mu_m = \frac{\sum_{k=1}^n \gamma_{m,k} \, p_k}{\sum_{k=1}^n \gamma_{m,k}},
\]
\[
\sigma_m^2 = \frac{\sum_{k=1}^n \gamma_{m,k} (p_k - \mu_m)^2}{\sum_{k=1}^n \gamma_{m,k}},
\]
\[
w_m = \frac{1}{n} \sum_{k=1}^n \gamma_{m,k}.
\]

In practice, the E- and M-steps are alternated until the bound on the data no longer increases and the parameters no longer change. This process iterates until convergence, even though the EM algorithm does not guarantee finding a global solution to this non-convex optimization problem. This way, the GMM model is well suited for this noise/defect classification task because it effectively captures the underlying structure of the data by representing it as a mixture of several Gaussian distributions. Each Gaussian component can model different characteristics of the data, allowing for the differentiation between noise and various types of defects based on their statistical properties.

However, to improve the robustness of the algorithm, we incorporate \textit{a priori} information on the model parameters, \( \Pr(\theta) \). This prior helps prevent scenarios where one of the Gaussian components becomes exclusively associated with a single data point. This situation can cause the variance of that component to gradually decrease, causing the likelihood to increase without bound \cite[p. 78]{Prince}. By introducing a prior distribution, we effectively regularize the model, ensuring that all components remain active and appropriately calibrated. This approach not only stabilizes the learning process but also improves the model's generalization capabilities by mitigating the risks of overfitting to noisy or anomalous data points. Consequently, the GMM is empowered to provide a more reliable classification between noise and defects, which ultimately leads to improved accuracy in image analysis tasks.

\begin{figure}[h!]
    \centering
    \includegraphics[width=\linewidth]{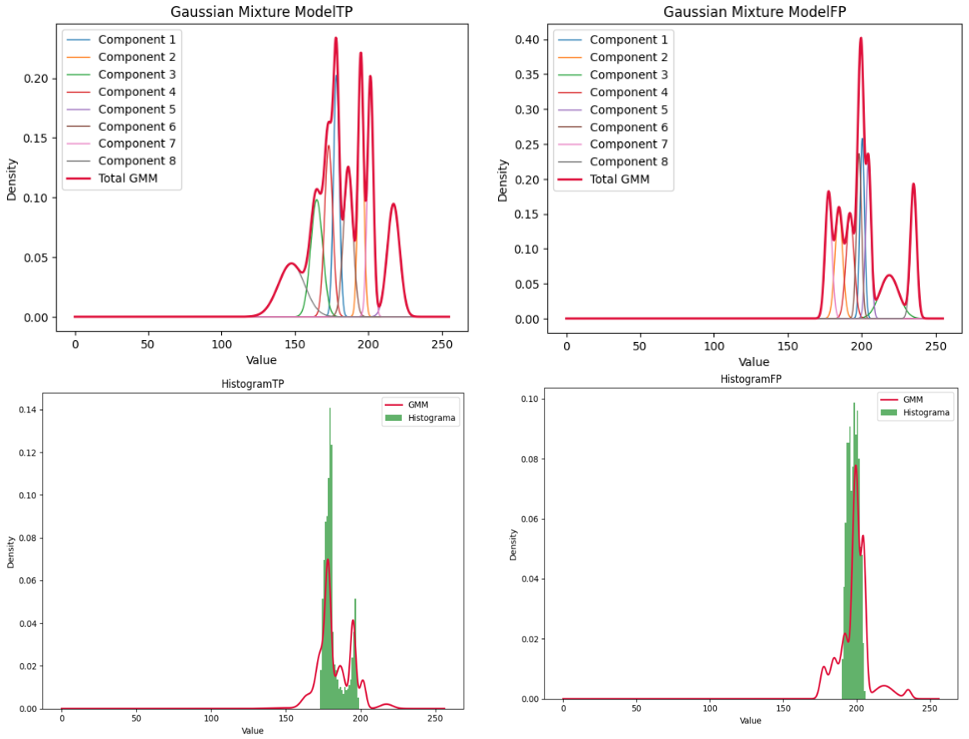}
    \caption{Two GMM fitted to FP and TP median histograms.}
    \label{fig:gmm_hist}
\end{figure}

\begin{figure}[h!]
    \centering
    \includegraphics[width=\linewidth]{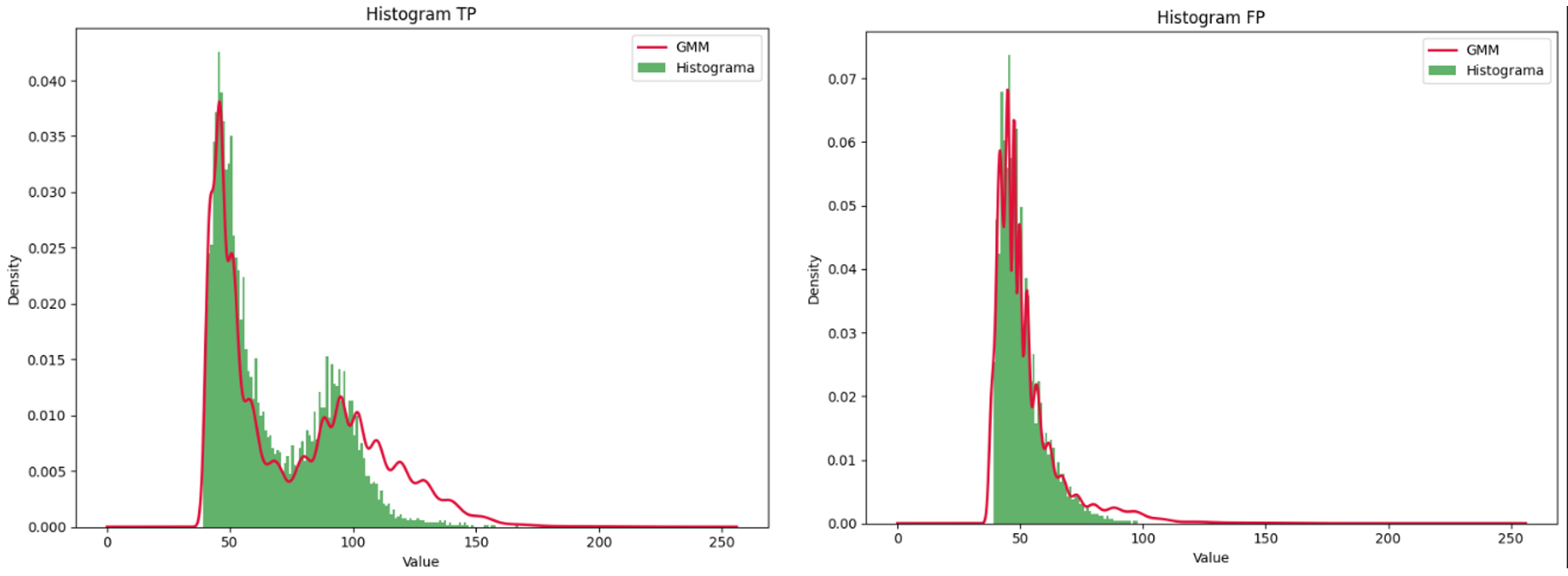}
    \caption{Visible singularity in defective histogram (left), compared to TP (right).}
    \label{fig:gmm_hist2}
\end{figure}

Once the GMM is fitted (see Fig.~\ref{fig:gmm_hist} and Fig.~\ref{fig:gmm_hist2}) to the histogram of \( u \), each pixel \( u_k \) can be assigned to the Gaussian component with the highest responsibility \( \gamma_{m,k} \), thus classified into the component \( k \) that maximizes responsibility:
\[
m_k^{\star} = \arg\max_m \{ \gamma_{m,k} \}.
\]
In this study, we adopt a post-processing approach to incorporate prior information into the GMM output. Based on empirical data and domain knowledge, we assume that approximately 35\% of the data corresponds to noise, while the remaining 65\% represents defects. To reflect this prior knowledge in our classification process, we define the prior probabilities \( \Pr(C_{\text{noise}}) = 0.35 \) and \( \Pr(C_{\text{defect}}) = 0.65 \).

After fitting the GMM, it provides the likelihoods \( \Pr(X \mid C_{\text{class}}) \) for each class, where \( C_{\text{class}} \in \{C_{\text{noise}} \equiv N, C_{\text{defect}} \equiv D\} \):
\[
\Pr(X \!\mid\! C_{\text{noise}}) = \sum_{k=1}^{K} \lambda_k \cdot \mathcal{N}_X(X \mid \mu_k, \Sigma_k),
\]
\[
\Pr(X \!\mid\! C_{\text{defect}}) = \sum_{k=1}^{K} \lambda_k \cdot \mathcal{N}_X(X \mid \mu_k, \Sigma_k).
\]

Once we have the likelihoods, we integrate the prior probabilities into the classification, updating them as follows:
\[
\Pr(X \!\mid\! C_{\text{noise}})_{\text{adjusted}} = \Pr(X \!\mid\! C_{\text{noise}}) \cdot \Pr(C_{\text{noise}}),
\]
\[
\Pr(X \!\mid\! C_{\text{k}})_{\text{adjusted}} = \Pr(X \!\mid\! C_{\text{k}}) \cdot \Pr(C_{\text{k}}).
\]

We can then apply Bayes' theorem to obtain the posterior probabilities:
\[
\Pr(C_{\text{noise}} \!\mid\! X) = \frac{\Pr(X \!\mid\! C_{\text{noise}})_{\text{adjusted}}}{\Pr(X)},
\]
\[
\Pr(C_{\text{defect}} \!\mid\!X) = \frac{\Pr(X \!\mid\! C_{\text{defect}})_{\text{adjusted}}}{\Pr(X)},
\]
where \( \Pr(X) \) is the marginal likelihood of the observed data \( X \) and represents the probability of observing the data regardless of the class to which it belongs (it serves as a normalization constant for the posterior probabilities). Beforehand, we compute \( \Pr(X) \) by summing the adjusted likelihoods:
\[
\Pr(X) = \sum_{C \in \{C_{\text{noise}}, C_{\text{defect}}\}} \Pr(X \!\mid\!C)_{\text{adjusted}}.
\]

This allows us to adjust the predicted probabilities based on our prior information. By combining the likelihood of the GMM with our heuristically derived priors, we effectively integrate this knowledge into the decision-making process. This approach improves the robustness of our classification task by ensuring that the inherent distribution of noise and defects is accurately represented in the model output. A threshold may be set on the posterior probabilities \( \gamma_{m,k} \) to classify pixels or neighborhoods as defective or nondefective based on the characteristics of the Gaussian components, such as casual pixel variations.

\subsection{Uniform Proximity for Pixel Value Distribution}
The reasoning for this metric is to think of the average proximity between pairs of points as a distinct distance measure that characterizes defects or noise. In a similar fashion to a previous feature, we perform a statistical test to extrapolate whether a patch is more likely to be a defect or noise on the basis of its spatial distribution.

We calculate the distance between points using the Euclidean, Manhattan, and Chebyshev distance:
\[
d_{\text{euclidean}} = \sqrt{(x_2 - x_1)^2 + (y_2 - y_1)^2},
\]
\[
d_{\text{manhattan}} = |x_2 - x_1| + |y_2 - y_1|,
\]
\[
d_{\text{chebyshev}} = \max\left(|x_2 - x_1|, |y_2 - y_1|\right).
\]

The average normalized distance between black pixels in an image is computed using:
\[
\bar{D} = \frac{1}{B \cdot M \cdot N} \sum_{i=1}^{B} d_{ij},
\]
where \( B \) is the number of black pixels and \( d_{ij} \) represents the distance between pair of pixels.

To evaluate the uniformity of the current candidate, we generate random noisy images that simulate the loss of black pixels. By creating multiple images with randomly distributed black pixels, we can compute the average distances between these pixels. This allows us to establish a reference distribution of distances, which serves as a benchmark to assess the uniformity of the candidate.

To determine whether a ROI is a defect or noise, we compute the p-value based on the actual mean distance \( D_{\text{actual}} \) and the simulated distances:
\[
p = \frac{1}{N} \sum_{i=1}^{N} \mathbf{1}_{(D_{\text{simulated},i} \geq D_{\text{actual}})}.
\]

\begin{figure}[h!]
    \centering
    \includegraphics[width=\linewidth]{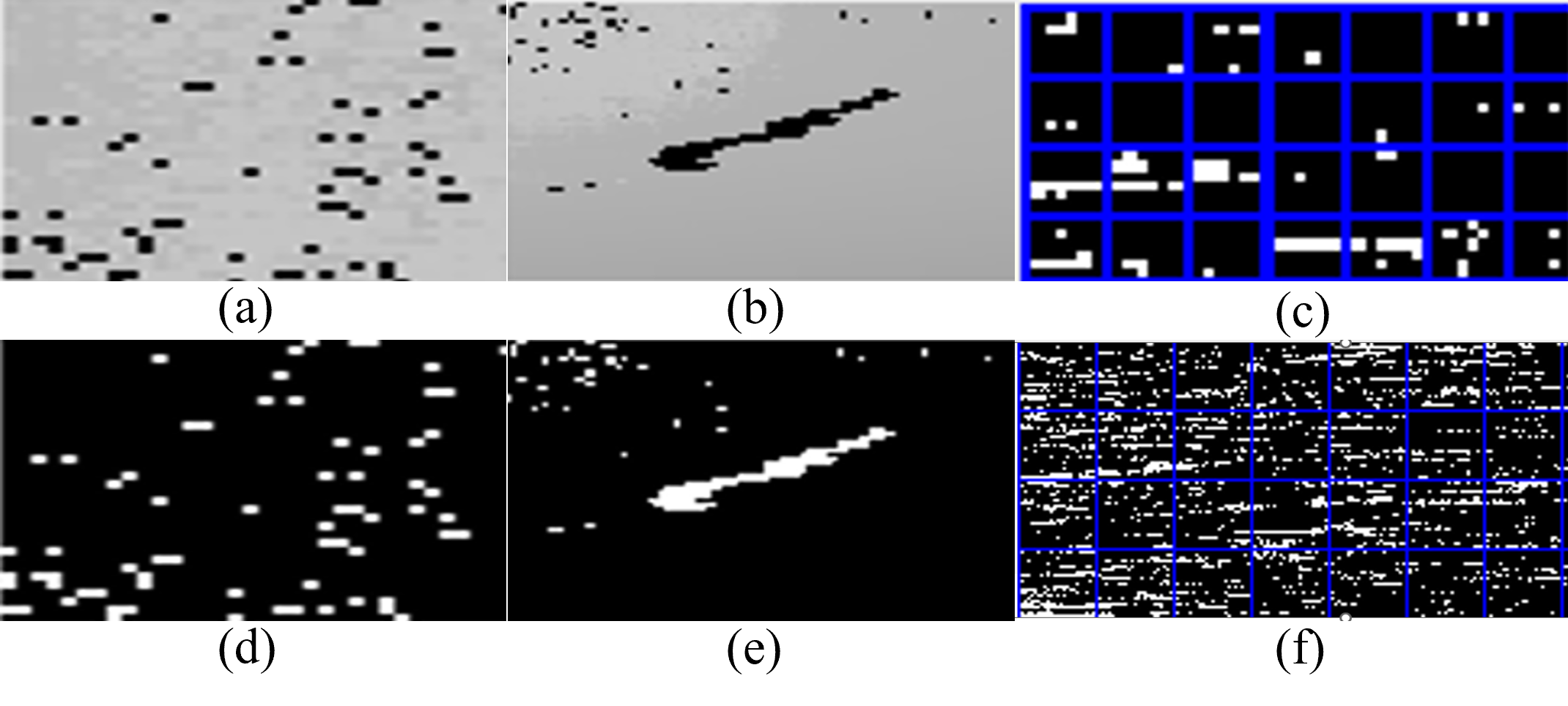}
    \caption{a, d) Clearly noisy regions. b, e) Clearly defective regions. c, f) Patch-based analysis of highly noisy regions.}
    \label{fig:examples}
\end{figure}

In Fig.~\ref{fig:examples}, examples of defective and noisy regions can be observed, and also two highly complex regions, characterized by a significant amount of noise mixed with defective areas caused by cracks and scratches.

\subsection{Connected Component Analysis for Structural Integrity}
A connected component analysis focuses on distinguishing the significant component of defects from noise by evaluating the number and size of connected components using 4- or 8-connectivity. The process involves isolating black pixels to calculate the ratio of isolated pixels and connected components. 

To improve the robustness of the ratio, we evaluated both the number and size of connected components. Let \( u \) be the binary image where \( u(x,y) \) with lost pixels \( u(x,y) = 0 \) representing potential defect areas. A connected component is defined as a cluster of connected pixels locatable by some labeling algorithm, following one of these approaches:
\begin{itemize}[label={-}]
    \item 4-Connectivity: Pixels \((x,y)\) are considered connected if they share an edge, which means that they are adjacent either horizontally or vertically, i.e. pixel \( k \) has a neighbor:
    \[
    \mathcal{N}_k = \{(x \pm 1, y), (x, y \pm 1)\}.
    \]
    \item 8-Connectivity: Pixels \((x,y)\) are considered connected if they share an edge or a corner, meaning they are adjacent either horizontally, vertically, or diagonally, i.e. pixel \( k \) has a neighbor:
    \[
    \mathcal{N}_k = \{(x \pm 1, y), (x, y \pm 1), (x \pm 1, y \pm 1)\}.
    \]
\end{itemize}

Once connected components are identified, we filter based on size. Let \( A(C_k) \) denote the area (number of pixels) of the connected component \( C_k \). To reduce the influence of noise, only components with an area larger than a threshold \( A_{\text{min}} \in \mathbb{N}^* \) are considered significant:
\[
C_k = \{C_k : A(C_k) \geq A_{\text{min}}\}.
\]

For each remaining connected component, we calculate a ratio that quantifies how isolated the lost pixels are within the component:
\[
R_{\text{isolated}} =  \frac{1}{M \cdot N} \frac{n_{\text{isolated}}}{n_{\text{components}}}.
\]

This involves counting how many black pixels are surrounded by white pixels within a chosen neighborhood \( \mathcal{N}_k \); \( n_{\text{isolated}} \) is the number of isolated pixels, \( n_{\text{components}} \) is the number of significant connected components and \( M \cdot N \) is the total number of pixels in the image. 

\subsection{Probabilistic Model Fitting and Feature Extraction for ROI}
The probability of observing this histogram, assuming that it follows a given distribution, is given as \( \Pr(x \!\mid\! \theta) \). The goal is to maximize the likelihood of observing this distribution given the histogram, \( \Pr(\theta \!\mid\! x) \), and determine which distribution is most likely to fit the histogram. The central premise is that the histograms of each class are likely to be distinguishable. By computing the intensity histograms for each image in a dataset and then averaging or taking the median across all images, we can create two global histograms: one for TP and one for FP. These global histograms are used as representative models for each class.

Given the differences in pixel distribution between TP and FP, particularly the fact that false positives often consist of many more black pixels (representing noise or defects), it is reasonable to hypothesize that the peak of the histograms will typically be concentrated near the lower intensity values. This observation helps guide our statistical modeling approach by providing a clear differentiation between the two classes and establishing a family of potential probability density functions (PDFs) to use.

We are interested in modeling the pixel intensity distribution as a probabilistic process. For each pixel \( x \), our objective is to find the most likely statistical distribution \( P(x \!\mid\! \theta) \), making the model parameters \( \theta \) contingent on the data. The key challenge is to maximize the likelihood of observing the actual pixel intensity values given the assumed distribution, or \( P(\theta \!\mid\! x) \), to determine which class (TP or FP) best fits the observed (current) histogram.

Let \( H_X \) be the histogram of the current image, and instead of averaging histograms for TP and FP, we define the median histograms \( H_{\text{TP}}^{\text{med}} \) and \( H_{\text{FP}}^{\text{med}} \) as follows:

For a set of histograms \( \{H_1, H_2, \ldots, H_k\} \) from a dataset of TP or FP cases, the median histogram is calculated bin-wise. For each bin \( i \), the median value across all histograms is given by:
\[
H^{\text{med}}(i) = \text{median}(H_1(i), H_2(i), \ldots, H_k(i)).
\]

\begin{figure}[h!]
    \centering
    \includegraphics[width=\linewidth]{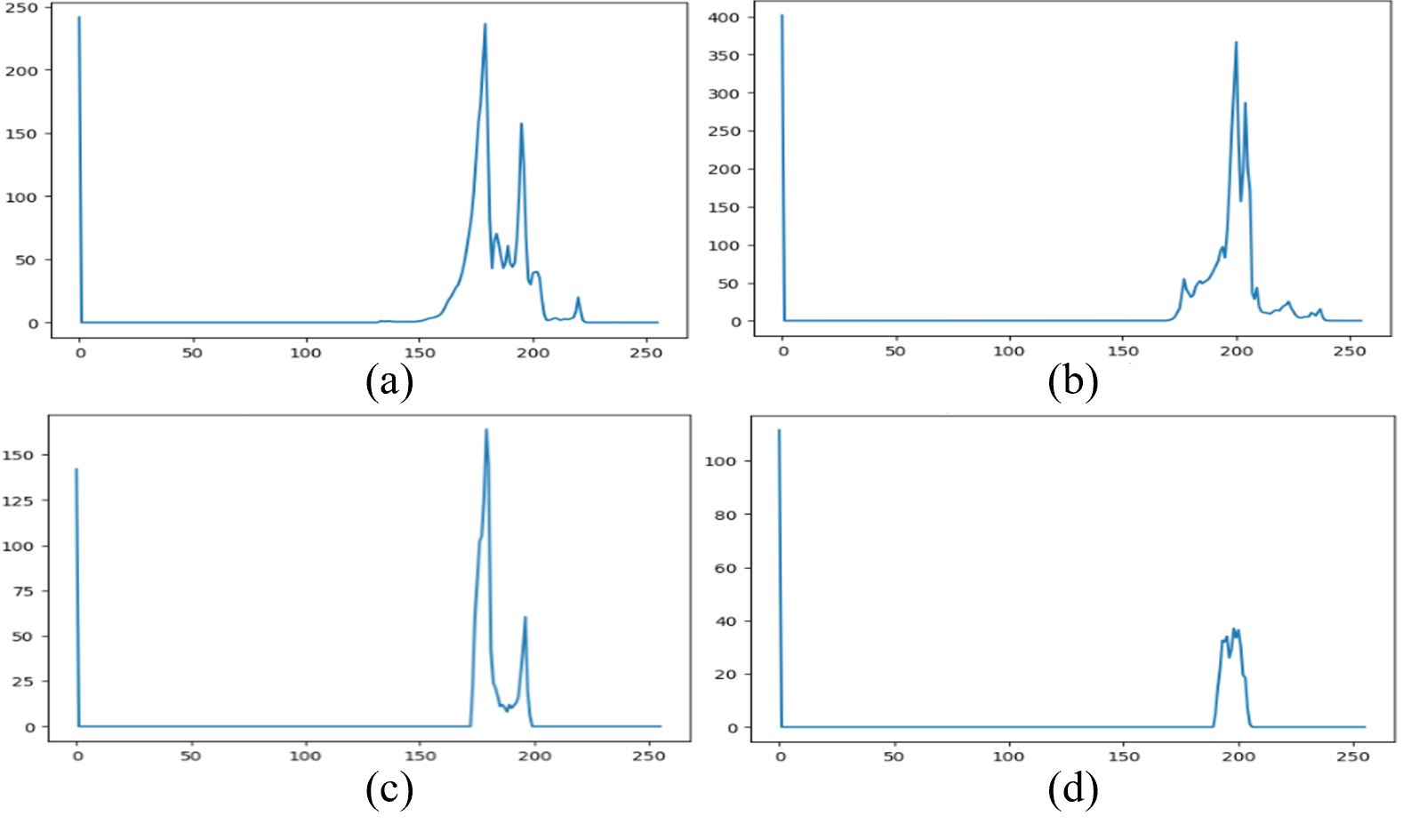}
    \caption{Comparison between (a, b) mean and (c, d) median histograms of FP and TP classes, respectively, for a basic features dataset.}
    \label{fig:med_hist}
\end{figure}

This produces the median histogram (see Fig.~\ref{fig:med_hist} for comparison), where each bin value \( H^{\text{med}}(i) \) represents the median of the corresponding bin values across the histograms in the dataset. 

The choice of using the median instead of the average provides several benefits. It is more robust against outliers, meaning that extreme values in certain bins of some histograms will not disproportionately influence the result. In addition, it ensures that the histogram reflects the “typical” structure of the dataset, reducing the effect of skewed or noisy data, which can occur when averaging over many histograms.

\begin{figure}[h!]
    \centering
    \includegraphics[width=\linewidth]{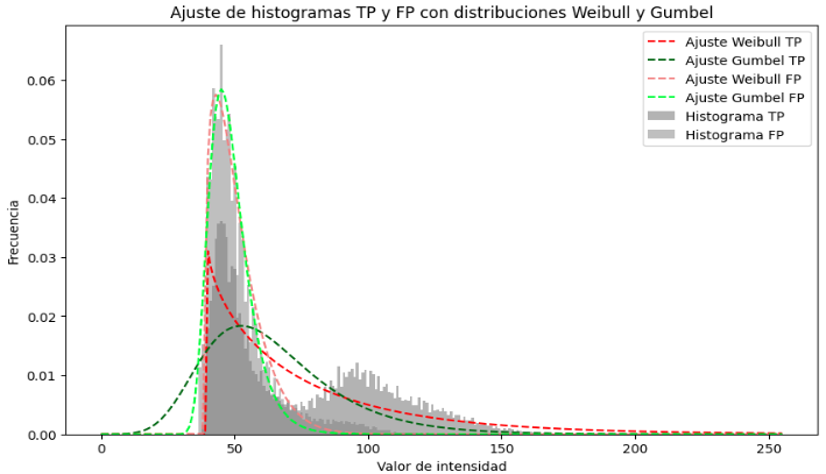}
    \caption{Different PDFs fitted to the classes' histograms. Red: Weibull TP fit, Blue: Gumbel TP fit, Orange: Weibull FP fit, Green: Gumbel FP fit, Light Gray: TP histogram, Dark Gray: FP histogram. Abscissa: intensity value, Ordinate: frequency. }
    \label{fig:gmm_hist3}
\end{figure}

We propose fitting the median histogram of each class to a GMM, a Weibull distribution, and a Gumbel distribution (see Fig.~\ref{fig:gmm_hist3}). The GMM assumes that the data is generated from a mixture of several Gaussian distributions, each component with its own weight \( w_i \), mean \( \mu_i \), and covariance matrix \( \Sigma_i \). Its likelihood is given by:
\[
\Pr(x \!\mid\! \theta) = \sum_{i=1}^{K} \left( w_i \cdot \mathcal{N}(x \!\mid\! \mu_i, \Sigma_i) \right).
\]

The Weibull distribution is often used to model failure rates and the distribution of extreme values in a dataset. Its PDF is given by:
\[
f(x; \lambda, k) = \frac{k}{\lambda} \left( \frac{x}{\lambda} \right)^{k-1} \exp\left(-\left(\frac{x}{\lambda}\right)^{k}\right),
\]
where \( k \) is the shape parameter and \( \lambda \) is the scale parameter. The Gumbel distribution is used to model the distribution of the maximum (or minimum) of a number of samples from a distribution. Its PDF is given by:
\[
f(x; \mu, \beta) = \frac{1}{\beta} \exp\left(-\frac{x - \mu}{\beta} - \exp\left(-\frac{x - \mu}{\beta}\right)\right),
\]
where \( \mu \) is the location parameter and \( \beta \) is the scale parameter. This distribution is often used to segment background and foreground due to its ability to model natural image histograms, as noted in \cite{Timm2011}.
In the test stage, we evaluated the likelihood of new image data (the pixel intensity values) under both the TP and FP fitted distributions, making the data contingent on the model parameters. To compare likelihoods, we use: \( \ell(x \!\mid\! \theta) = \log(\Pr(x \!\mid\! \theta))\) and we normalize the pixel-wise log-likelihoods to determine which model has the highest logarithmic probability for more pixels:
\[
\ell_{\text{norm,TP}}(x_i) = \frac{\ell(x_i \!\mid\! \theta_{\text{TP}})}{\sum_j \ell(x_j \!\mid\! \theta_{\text{TP}})},
\]
\[
\ell_{\text{norm,FP}}(x_i) = \frac{\ell(x_i \!\mid\! \theta_{\text{FP}})}{\sum_j \ell(x_j \!\mid\! \theta_{\text{FP}})},
\]
\[
N_{TP} = \sum_i \mathbf{1}_{\left(\ell_{\text{norm,TP}}(x_i) > \ell_{\text{norm,FP}}(x_i)\right)}.
\]

In addition, to provide a more robust comparison that preserves the sign of the likelihood values, we calculate the sum of the log-likelihoods raised to the third power:
\[
S_{\text{TP}} = \sum_i \ell(x_i \!\mid\! \theta_{\text{TP}})^3,
\]
\[
S_{\text{FP}} = \sum_i \ell(x_i \!\mid\! \theta_{\text{FP}})^3.
\]

\subsection{Centroid Proximity Filtering for Candidate Selection}
This metric involves centroid proximity filtering to select candidate regions with potential defects. The centroid \((x_c, y_c)\) of the black pixels is given by:
\[
x_c = \frac{1}{P_{\text{black}}} \sum_{x=1}^{H} \sum_{y=1}^{W} x \cdot \mathbf{1}_{(u(x,y) = 0)}, 
\]
\[
y_c = \frac{1}{P_{\text{black}}} \sum_{x=1}^{H} \sum_{y=1}^{W} y \cdot \mathbf{1}_{(u(x,y) = 0)}.
\]

The centroid represents the geometric center of the black pixels; proximity to this centroid can be used to filter out regions that do not exhibit defect-like behavior.

\subsection{Spatial Clustering Analysis of Pixel Value Distribution}
This metric is based on the total variation (TV) and the integral of spatially oriented sums of pixels, applied to pixels lost in the image. The key assumption is that a defect becomes more evident when lost (e.g. black) pixels cluster together, so dispersion of black pixels would indicate less defined defects, whereas higher agglomeration (concentration) would signal more prominent defects.

In our problem, due to the nature of the sensor used, the distribution of black pixels is essential for determining clustering or agglomeration, and it can be assessed by calculating the sum of black pixels across different axes, including rows, columns, and diagonals.

The number of black pixels, denoted as \( B \), is defined as
\[
B = \sum_{x=1}^{H} \sum_{y=1}^{W} \mathbf{1}_{(u(x,y) = 0)}.
\]

We compute three key sums of black pixels:
\begin{itemize}[label={-}]
    \item Row Sum: The sum of black pixels across each row \( i \):
    \[
    R(i) = \sum_{j=1}^{W} \mathbf{1}_{(u(i,j) = 0)}.
    \]
    \item Column Sum: The sum of black pixels across each column \( j \):
    \[
    C(j) = \sum_{i=1}^{H} \mathbf{1}_{(u(i,j) = 0)}.
    \]
    \item Diagonal Sum: The sum of black pixels along each diagonal \( d \), which includes diagonals that traverse both the top left to the bottom right and the top right to the bottom left:
    \[
    D(d) = \sum_{(i,j) \in \text{diagonal } d} \mathbf{1}_{(u(i,j) = 0)}.
    \]
\end{itemize}

TV measures the rate of change in the distribution of black pixels and can be used to quantify the clustering of black pixels. In a discrete setting it is calculated as:
\[
TV(f) = \sum_{i=1}^{N} \left| f(i+1) - f(i) \right|,
\]
with \( f : \mathbb{N} \to \mathbb{R} \) and \( f(i) \in \{R(i), C(i), D(i)\} \). The total variation captures the degree of change between neighboring rows, columns, or diagonals. 

We then normalize TV with respect to the number of black pixels:
\[
\overline{TV}(f) = \frac{TV(f)}{P_{\text{black}}}.
\]
This normalization helps ensure that the TV metric is scale-independent and reflects the relative distribution of black pixels.

The integral of the discrete 1D signal (black-pixel oriented sums) measures the overall accumulation of black pixels in the region. For a signal \( f(i) \), the normalized integral is computed as:
\[
\overline{\text{Int}}_f = \frac{1}{N} \sum_{i=1}^{N} f(i),
\]
where \( N \) is the length of the signal (number of rows, columns, or diagonals).

To further enhance defect detection, a combination of Total Variation and integral metrics is proposed. The overall joint defect metric \( D_{\text{combined}} \) is calculated as:
\[
D_{\text{combined}} = \alpha \cdot \overline{TV}_{\text{combined}} + \beta \cdot \overline{\text{Int}}_{\text{combined}}
\]
\[
= \sqrt{\overline{TV}_{\text{row}}^2 + \overline{TV}_{\text{col}}^2 + \overline{TV}_{\text{diag}}^2} \hspace{0.1cm} + 
\]
\[
\sqrt{
\left(\overline{\text{Int}}_{\text{row}} f \right)^2 + 
\left(\overline{\text{Int}}_{\text{col}} f \right)^2 + 
\left(\overline{\text{Int}}_{\text{diag}} f \right)^2
},
\]
where \( \alpha \) and \( \beta \) are real weighting factors that balance the contributions of TV and the integral. 

\begin{figure}[h!]
    \centering
    \includegraphics[width=\linewidth]{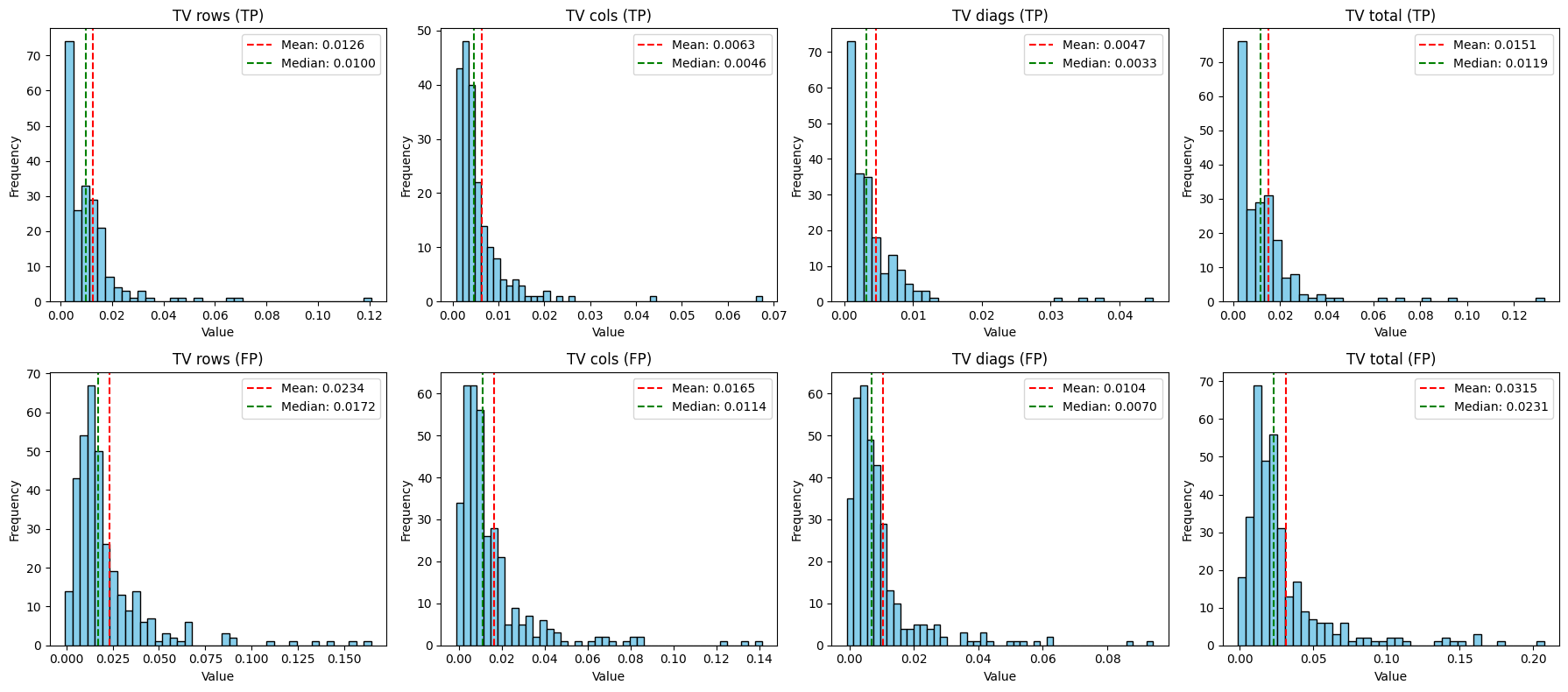}
    \caption{Histograms of the TV feature for FP and TP data.}
    \label{fig:tv_mean_median}
\end{figure}

\begin{figure}[h!]
    \centering
    \includegraphics[width=\linewidth]{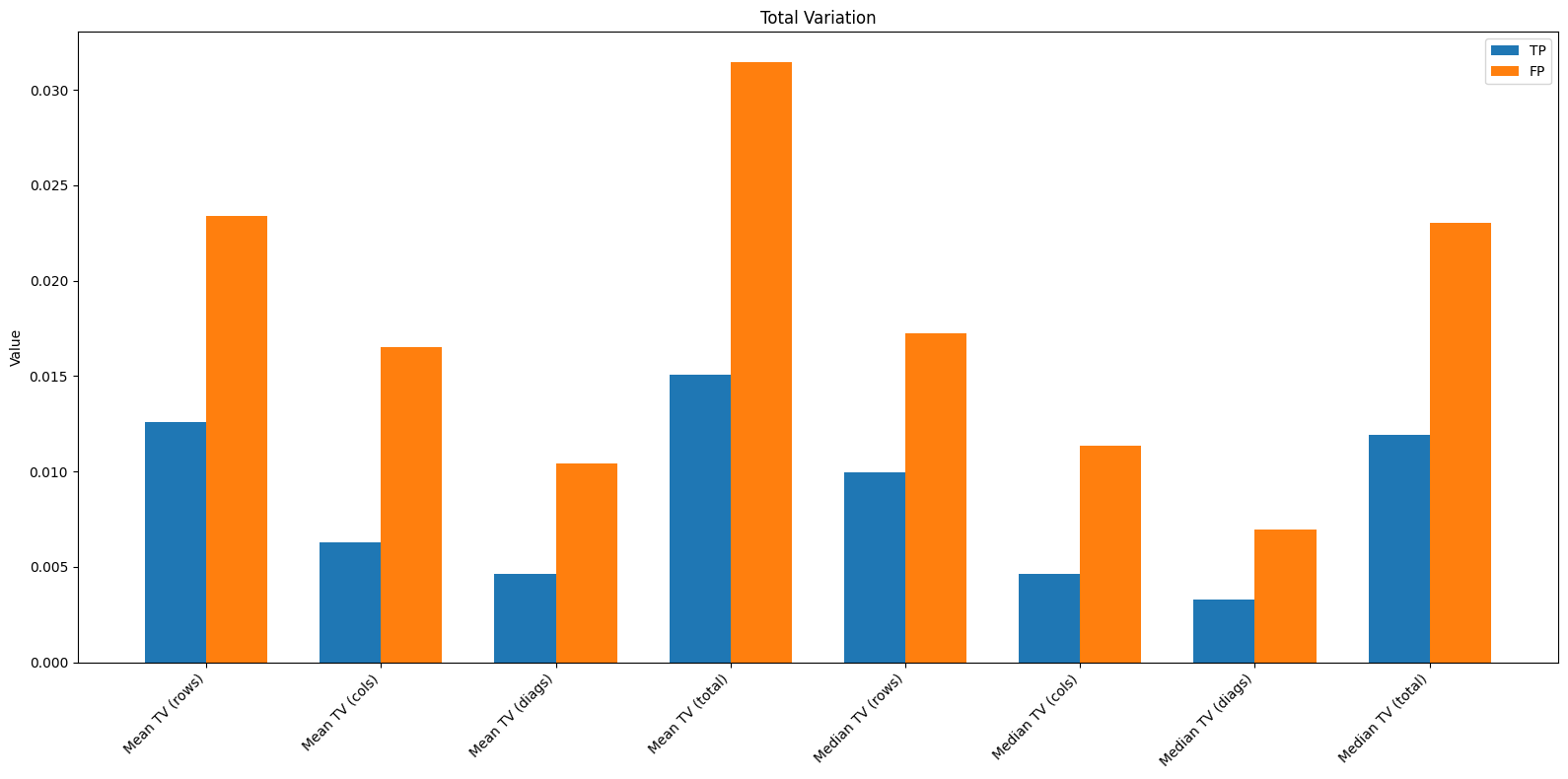}
    \caption{Comparison of the TV feature for FP and TP data.}
    \label{fig:tv}
\end{figure}

Upon applying these metrics to the dataset, we can compute the mean and median TV for TP and FP, as well as the mean and median integral values for TP and FP. The results indicate that images with genuine defects (TP) tend to have higher values of both TV and integral, confirming the clustering of black pixels around defect regions. See Fig.~\ref{fig:tv_mean_median} and Fig.~\ref{fig:tv} for some quantitative results obtained in the classification capabilities based on the TV feature. The separation between classes is remarkably good.

\subsection{Distance Metrics for Proximity Analysis of Detection Points}
To assess the similarity between the current image histogram and the median histograms of TP and FP, several comparison metrics can be employed. In this context, we propose the utilization of metrics like the Hellinger/Bhattacharyya distance, chi-square distance, correlation, and intersection. The response (measure of similarity between two histograms) of these metrics differs; we now define each of them:

\begin{itemize}[label={-}]
    \item {Bhattacharyya Distance (or Hellinger Distance):} This metric measures the overlap between two histograms. A lower value indicates greater similarity between the two histograms.
    \[
    d_{\text{B}}(H_1, H_2) = \sqrt{1 - \frac{1}{\sqrt{\bar{H}_1 \bar{H}_2 N^2}} \sum_{i=1}^n \sqrt{H_1(i) \cdot H_2(i)}},
    \]
    where \(\bar{H}_1\) and \(\bar{H}_2\) are the mean values of the histograms \(H_1\) and \(H_2\):
    \[
    \bar{H}_k = \frac{1}{n} \sum_{i=1}^n H_k(i), \quad k \in \{1, 2\}.
    \]

    \item {Correlation Metric:} This metric measures the linear relationship between the two histograms. A higher value indicates a stronger correlation:
    \[
    \hspace*{-\leftmargin} d_{\text{Corr}}(H_1, H_2) = \hspace{4.75cm}
    \]
    \[
    \frac{\sum_{i=1}^n (H_1(i) - \bar{H}_1)(H_2(i) - \bar{H}_2)}{\sqrt{\sum_{i=1}^n (H_1(i) - \bar{H}_1)^2 \sum_{i=1}^n (H_2(i) - \bar{H}_2)^2}}.
    \]

    \item {Intersection Metric:} This metric measures the amount of overlap between two histograms. A higher value indicates greater overlap and therefore greater similarity:
    \[
    d_{\text{Intersection}}(H_1, H_2) = \sum_{i=1}^n \min(H_1(i), H_2(i)).
    \]
\end{itemize}

First, histograms are extracted from the dataset for both TP and FP. For each of these, instead of calculating the average, the median is computed bin-by-bin, as described earlier.

During testing, the histogram of the current image, denoted as \( H_X \), is compared against the median histograms \( H_{\text{TP}}^{\text{med}} \) and \( H_{\text{FP}}^{\text{med}} \) using various similarity metrics. When evaluating the metrics for the current histogram against only true positives (i.e. true defects), the Bhattacharyya/Hellinger distance produced the most accurate results with minimal variation across the TP histograms, indicating that these metrics are robust in recognizing true positives.

Similarly, when evaluating the current histogram against only false positives (i.e., true noise), the results showed that the Bhattacharyya/Hellinger distance consistently provided the lowest false positive detection rates, making it the most reliable for distinguishing between TP and FP cases. Given its stability and reliability in both cases, we utilize this metric as a primary approach for classification.

\subsection{Basic features}
We finalize this section by introducing a set of proposed analytical and very straightforward metrics designed to address the classification problem by leveraging various image characteristics from different semantic perspectives, including textural, structural, spatial, and spectral features. These metrics differ from the previously introduced ones in that they offer a higher-level rationalization and are more direct in their approach. While the earlier metrics are built on deeper, more intuitive foundations, these new metrics take a more generalized approach. However, they still maintain a solid theoretical basis for their use as feature extractors.

\begin{figure}[h!]
    \centering
    \includegraphics[width=\linewidth]{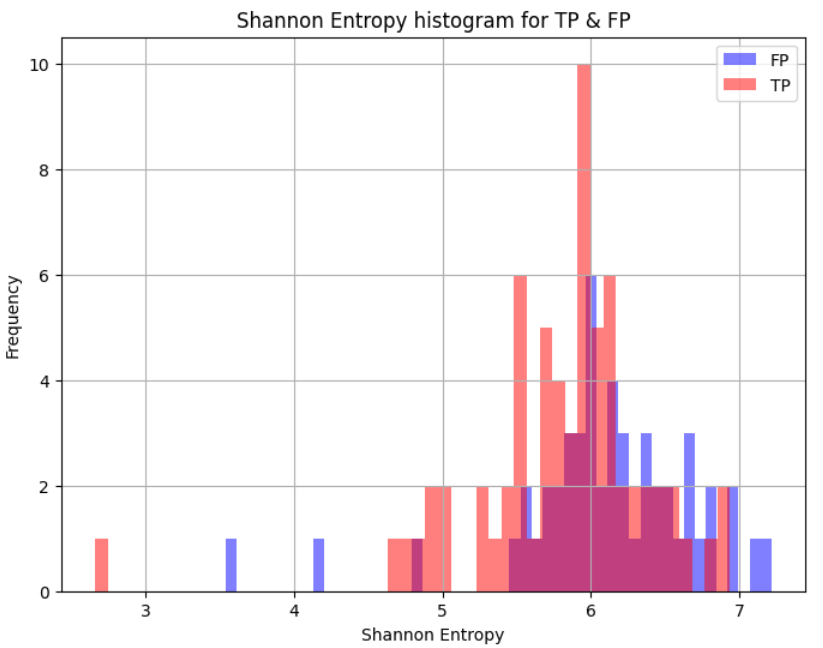}
    \caption{Feature with poor inter-class separability.}
    \label{fig:poor}
\end{figure}

\begin{figure}[h!]
    \centering
    \includegraphics[width=\linewidth]{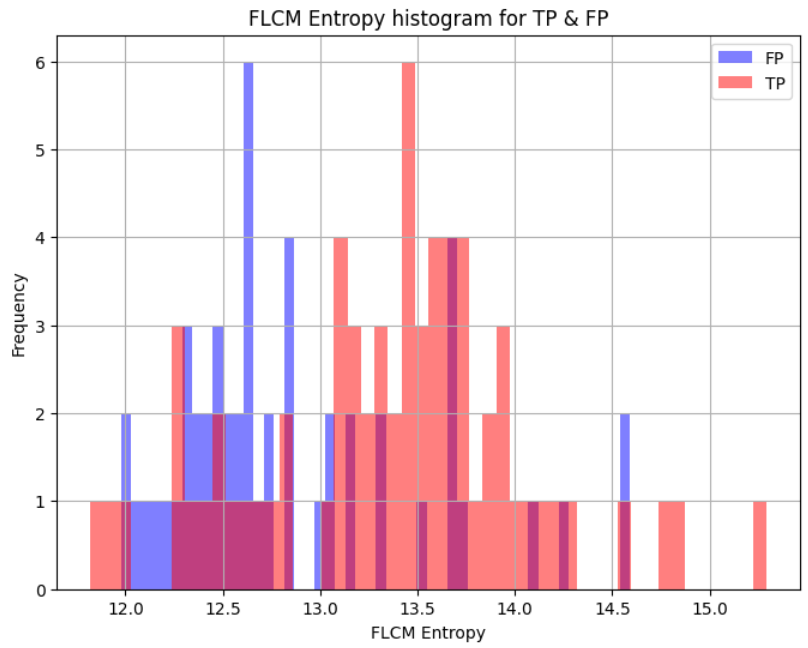}
    \caption{Feature with weak inter-class separability.}
    \label{fig:weak}
\end{figure}

An ideal scenario involves identifying two fundamental features that can satisfactorily separate the clusters of each class, either linearly or polynomially. This can be explored at a basic level using data visualization or pattern recognition techniques. For example, consider Figs.~\ref{fig:poor} and \ref{fig:weak}, which illustrate two cases: one where the feature is poorly effective and the other where it performs slightly better. This is evaluated on the basis of how well the mean or median of the histograms are separated.
In general, to ensure consistency across images from different sources, we propose normalizing them to a common reference. This can be achieved without degrading information by applying dynamic range-based intensity normalization, using methods such as Z-score normalization and linear interpolation, to transform pixel intensity values into a standardized range of \([0, L-1)\), where \(L = 2^b\) represents the number of discrete intensity levels. Here, \(b\) is typically 8 for 8-bit images or 16 for 16-bit images, corresponding to powers of two.

For an image \(u(x, y)\), we define the Z-score normalization as:
\[
u_{\text{z-norm}}(x, y) = \frac{u(x, y) - \mu_u}{\sigma_u},
\]
where \( \mu_u \) represents the normalized value of the pixel at
position \((x, y)\) and \( \sigma_u \) is the standard deviation of the intensity values. This normalization ensures that the image data have a mean of 0 and a standard deviation of 1, allowing for consistent feature extraction across varying sources and conditions.

In the following sub-subsections, we briefly introduce the methods categorized in this group:

\subsubsection{Point loss} 
\hspace*{0.25cm} The laser triangulation sensor used returns floating point zeros for points where measurements cannot be obtained. These zeros manifest as null values in the distance or intensity images, but appear as indeterminate extreme values in the normal maps. Therefore, if there is significant point loss within the bounding box of a candidate, as reflected in the distance map (i.e., many null values), the analysis of that candidate is halted immediately, and it is labeled accordingly.

The algorithm evaluates a candidate by comparing histograms and correlations between neighboring regions.

Initially, a central rectangle is defined based on the candidate's coordinates, and, for efficiency, neighboring rectangles are created adjacent to each edge of the central rectangle. Consequently, the candidate's neighborhood is approximated using its four rectangular neighbors, maintaining the same aspect ratio unless they exceed the image boundaries, in which case their sizes are maximized based on the upper limits.

Alternatively, the 2D spatial distribution of the points lost within the candidate is approximated by calculating the mean and variance of the coordinates \( \{(x,y) \!\mid\! u(x,y)=0\} \), followed by determining the standard deviation in both the \( x \) and \( y \) directions. If both standard deviations are below a predefined threshold, it indicates that the pixels are not clustered under a normal distribution at a level of \( 3\sigma \). This suggests that the distribution of the black pixels is nonclustered, implying that the candidate is likely to represent a TP.

Conversely, if either of the standard deviations exceeds the threshold, it implies that the distribution of the black pixels is clustered, which may indicate the presence of random noise or a FP. The decision-making aspect is reserved for the machine learning model, while we focus on calculating the \( l^2 \)-norm of the deviations.

From here, \textit{one} characteristic emerges.

\subsubsection{SSIM} 
\hspace*{0.25cm} This algorithm calculates the Structural Similarity Index (SSIM) between the candidate and its four neighboring regions. SSIM is a measure of similarity between two images at different levels, considering luminance, contrast, and structure (shape). The objective is to determine whether any of the neighbors is very similar to the candidate, which could indicate the presence of false positives (FP).

SSIM captures the interdependence between pixels, particularly in their neighborhoods, quantifying how distortions in an image are less perceptible in brighter areas and when textures are present. SSIM is calculated using the multiplicative combination of three components: luminance \( l(X,Y) \), contrast \( c(X,Y) \), and structural information \( s(X,Y) \). These components are defined mathematically as follows:
\[
l(X,Y) = \frac{2 \mu_X \mu_Y + C_1}{\mu_X^2 + \mu_Y^2 + C_1}, \
c(X,Y) = \frac{2 \sigma_X \sigma_Y + C_2}{\sigma_X^2 + \sigma_Y^2 + C_2}, 
\]
\[
s(X,Y) = \frac{\sigma_{XY} + C_3}{\sigma_X \sigma_Y + C_3},
\]
where \( \mu_X \) and \( \mu_Y \) are the means of the images, \( \sigma_X \) and \( \sigma_Y \) are their standard deviations, \( \sigma_{XY} \) is their covariance, and \( C_1 \), \( C_2 \), and \( C_3 \) are small constants used to prevent division by zero, parametrized by 0. The general SSIM index is given by:
\[
SSIM(X,Y) = l(X,Y)^\alpha \cdot c(X,Y)^\beta \cdot s(X,Y)^\gamma,
\]
with default values set for \( \alpha \), \( \beta \), and \( \gamma \) to 1. The SSIM values are computed through Gaussian convolution of the images to derive their means and variances, ultimately resulting in a scalar value representing the structural similarity between the candidate patches. Although not specifically designed for object identification, low SSIM values can indicate potential problems in the frame, such as excessive noise.

\subsubsection{Basic Statistical Metrics}\label{sec:basic_statistical_metrics} 
\hspace*{0.25cm} The dynamic range of the candidate image provides a comprehensive measure of brightness, contrast, and potential noise. Key statistical indicators include the average (mean intensity), mode (most frequent), median (robust central value), maximum (brightest, indicating possible overexposure), and variance \((\sigma^2)\) (contrast, spread, or dispersion) of the intensity value. These metrics, when analyzed together, reveal both general brightness and noise presence in the image:
\[
\text{Average} = \frac{1}{N} \sum_{i=1}^{N} I(i), \quad \text{Max} = \max(I(i)),
\]
\[
\sigma^2 = \frac{1}{MN} \sum_{z=0}^{M-1} \sum_{y=0}^{N-1} (I(x, y) - \mu)^2,
\]
\[
\text{Median} = 
\begin{cases} 
    x_{\text{ordered}\,(N+1)/2}, & N \text{ odd} \\
    \frac{x_{\text{ordered}\,N/2} + x_{\text{ordered}\,(N/2+1)}}{2}, & N \text{ even}.
\end{cases}
\]

A low dynamic range, calculated as \(\text{Dynamic Range} = \max(I(i)) - \min(I(i))\), may signal poor feature extraction potential due to lack of variability, leading to areas filled with lost points. Additionally, in highly homogeneous regions, defects with gray levels significantly distinct from their surroundings could justify retaining the candidate as a true positive (TP), reducing misclassification risk.

Although simple in nature, these features capture noise distribution or data loss compared to other metrics, as they represent the simplest characteristics expected to hold information about the nature of the noise. These include variance, covariance, mode, median, coefficient of variation, and covariance ratio, each providing unique insights into the image's statistical properties.

The covariance between two random variables in the image describes the degree to which these variables vary together. It is calculated by treating each column of the candidate image as a variable and each row as a sample:
\[
\text{Cov}(X, Y) = \frac{1}{N} \sum_{z=0}^{N} (x_i - \mu_X) \cdot (y_i - \mu_Y).
\]
The determinant of the covariance matrix can be used to assess variability between multiple variables.

The coefficient of variation (CV) is defined as the ratio of the standard deviation to the mean, provides a normalized measure of variability, making it easier to compare images with different average intensities:
\[
\text{CV} = \frac{\sigma}{\mu} = \frac{\sqrt{\sigma^2}}{\mu}.
\]

Finally, the covariance ratio at low frequencies (CRLF) compares the covariance of the original image \( I \) with that of a Laplacian-filtered (high-pass) version:
\[
\text{CRLF} = \frac{\text{Cov}_{I}}{\text{Cov}_{I * L}},
\]
where \( L \) is a discretized Laplacian kernel applied using convolution to highlight high-frequency features (edges) in the image.

A unitary positive ratio suggests that the filtered candidate's pixels tend to vary together, indicating a plausible FP.

From here, eight characteristics emerge.

\subsubsection{Texture metrics} 
\hspace*{0.25cm} First, several metrics are calculated to capture information about the texture of the candidate, specifically regarding the distribution of pixel intensities. Initially, the texture characteristics of all candidates from each image are calculated and stored separately to enable global normalization. 

The following equations explain the characteristics implemented that encapsulate information about the texture of the candidate. Since they depend on the pixel values of the candidate region, it becomes necessary to normalize them both initially (first normalization) and subsequently (both locally and globally).

\begin{itemize}[label={-}]
    \item {Entropy}: Using Shannon's formula, where \( p \) is the probability of occurrence of the gray level \( x_i \) (i.e., the normalized histogram value at that level):
    \[
    H(X) = - \sum_{i=0}^{n-1} p(x_i) \cdot \log_2(p(x_i)).
    \]
    
    \item {Energy}: The sum of the squares of the intensity values of all pixels:
    \[
    E = \sum_{x=0}^{M-1} \sum_{y=0}^{N-1} I(x, y)^2.
    \]
    
    \item {Homogeneity}: The sum of the probabilities of the occurrence of gray values weighted by their distance to the mean value (with \( p \) being the normalized histogram vector as the evaluation is based on the histogram vector, not all candidate pixels):
    \[
    \text{Homogeneity} = \sum_{i=0}^{n-1} \frac{1}{1 + |i - j|} p(i, j).
    \]

    \item {Contrast}: The standard deviation of the intensity of the candidate's pixels.
    
    \item {Kurtosis}: The normalized fourth central moment divided by the fourth power of the candidate's standard deviation:
    \[
    K = \frac{\mu_4}{\sigma^4} = \frac{1}{N} \sum_{x} \sum_{y} (x - \bar{x})^4 f(x, y).
    \]
\end{itemize}

Other metrics, such as skewness and variance, may be redundant, and only these are implemented. Although it may seem that many metrics would introduce redundancy—and in some cases, they would—there may be instances where the evolution of one is more closely related to the values of the random variable than the others. Therefore, all metrics are evaluated with each training dataset.

Next, a ``local'' normalization of these characteristics is performed. Each characteristic is adjusted by subtracting the mean of the image and dividing by its standard deviation. Finally, the values that will persist in the feature database undergo a ``global'' normalization, utilizing the metrics of all candidates in each sample (total normal image). This process scales the dynamic range to the minimum and maximum of each characteristic, resulting in a unit norm (converting them to a range [0, 1]).

From here, six characteristics emerge.

\subsubsection{Gabor filter in YUV} 
Gabor filters are used to capture texture information in an image by highlighting specific spatial frequencies and orientations. In this approach, we apply Gabor filters, i.e. Fourier basis functions modulated by Gaussians across different directions, to extract two key statistical features: energy and variance. These filters, with real and imaginary components, detect structures at various angles and frequencies, allowing robust texture analysis.

Instead of using an RBG image, one could better focus on the luminance channel (Y$'$ or Y) from the YUV color space, which isolates brightness variations, as chrominance (color information) is not as useful in this particular setting. The real and imaginary components of the Gabor filter correspond to orthogonal directions, and the Gabor filter itself can be expressed in its complex form as:
\[
G(x, y) = \exp \left( -\frac{1}{2} \left( \frac{x'^2}{\sigma_x^2} + \frac{\gamma^2y'^2}{\sigma_y^2} \right) \right) \cos \left( 2 \pi \frac{x'}{\lambda} + \psi \right),
\]
where \(\lambda\) represents the wavelength of the sinusoidal factor, \(\sigma_x\) and \(\sigma_y\) are the standard deviations of the Gaussian in the \(x\)- and \(y\)-directions, respectively, and \(\gamma\) is the aspect ratio controlling the elliptical shape of the filter. \(f\) is the frequency of the sinusoidal wave and \(\psi\) is the phase offset. We have only used the real part of the Gabor filter bank; as indicated in \cite{Jing2014} it is sufficient for our defect detection tasks.

\begin{figure}[h!]
    \centering
    \includegraphics[width=\linewidth]{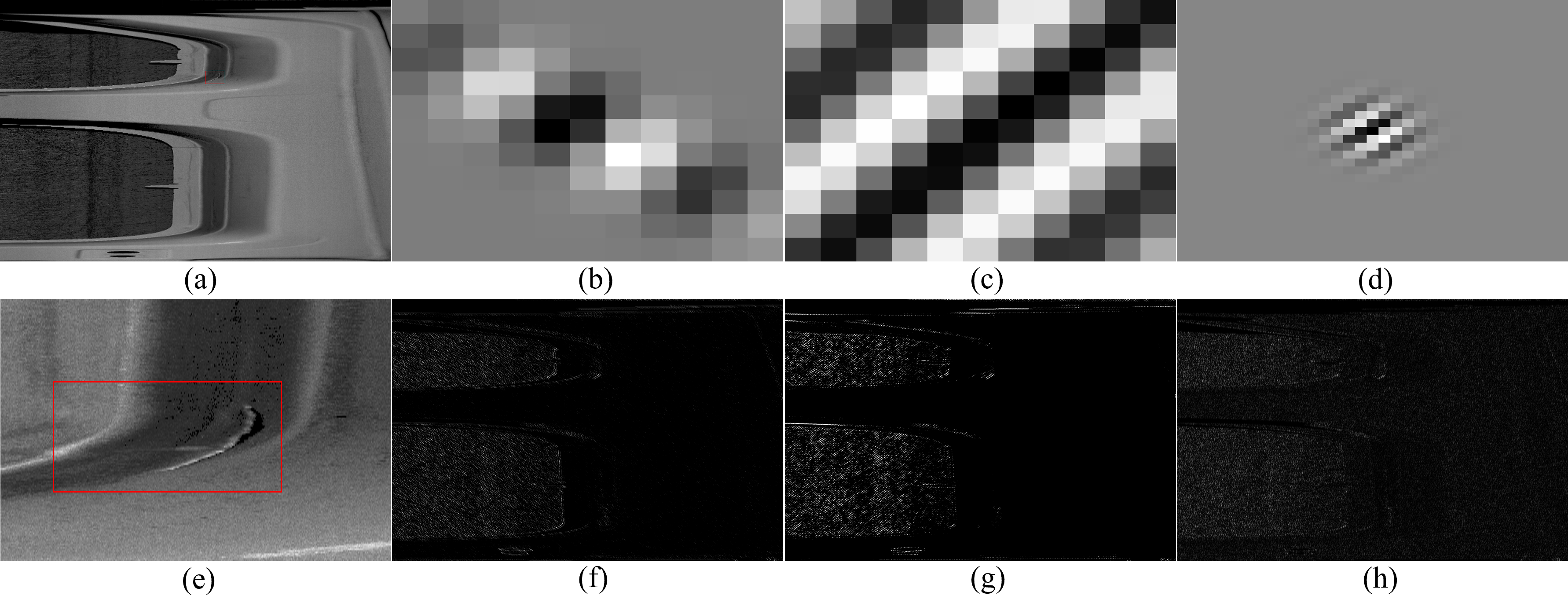}
    \caption{a, b) Defective candidate and defect's ROI. b-d) Three different Gabor filters. f-h) Three different spatio-temporal responses to the Gabor filters.}
    \label{fig:gabor}
\end{figure}

In our method, we apply Gabor filters in four orientations (every \(45^\circ\)), then accumulate the energy from the filtered images and normalize it to the dynamic range of the image. This process enhances the visibility of defect details at specific spatial frequencies and orientations (see Fig.~\ref{fig:gabor}). The accumulated responses are also used to compute the variance, which helps to distinguish the pattern of defects. This multidirectional filtering allows us to detect textual anomalies while avoiding the diffusion of detail, leading to sharper defect detection.
This method proves particularly useful for identifying defects like surface dents, as it emphasizes directional features and captures texture-specific anomalies in the image.

From here, two characteristics emerge.

\subsubsection{Homomorphic filter} 
\hspace*{0.25cm} The homomorphic filter is employed to enhance image contrast by suppressing low frequencies—corresponding to large, uniform areas of intensity—while amplifying higher frequencies to capture finer details. It is especially useful for restoring degraded images with uneven lighting and intensity, as it sharpens edges more effectively than a conventional high-pass filter. By doing so, it highlights small-scale features and outlines of potential defects, while flattening homogeneous regions.

An image \( I(x, y) \) can be represented as the product of two components: illumination \( L(x, y) \) and reflectance \( R(x, y) \). Then one can take a logarithmic transformation to turn the multiplicative model into an addition and by means of the Discrete Fourier Transform (DFT) apply a high-pass filter in the frequency domain to the log-transformed, before taking the inverse Fourier transform and exponentiation are performed to revert to the spatial domain:
\[
\log(I(x, y)) = \log(L(x, y)) + \log(R(x, y)),
\]
\[
F(u, v) = H(u, v) \cdot \mathcal{F} \left[ \log(I(x, y)) \right],
\]
\[
I(x, y) = \exp \left( \mathcal{F}^{-1} \left( \mathcal{F} \left[ \log(I(x, y)) \right] \right) \right).
\]

This filtering technique helps suppress frequencies often associated with large, uniform areas of intensity (e.g. shadows), at the same time carefully enhancing higher frequencies (details) and improving visibility of significant features, addressing issues related to illumination and contrast irregularities.

The key assumption is that the defect's shape is more distinct and irregular than its surroundings, and its contours will exhibit unpredictable variations compared to nearby regions. To quantify this irregularity, the entropy of the region is calculated. Entropy measures the amount of information or disorder present in the high-frequency components of the candidate, giving an indication of how chaotic or structured the region is:
\[
H = - \sum_{i} p_i \log(p_i),
\]
where \( p_i \) is the probability of occurrence of the intensity level in the ROI.

Using this method, defect patterns are made more prominent, as homomorphic filtering focuses on variations in texture and contour, helping to isolate defects with distinct edges from otherwise smooth areas.

From here, one characteristic emerges.

\subsubsection{HOG vector field} 
\hspace*{0.25cm} The Histogram of Oriented Gradients (HOG) descriptor is calculated with a focus on generalizing various types of defects. Each orientation calculated by the HOG descriptor is represented as a complex number with its x and y components, allowing the candidate image to be understood as a vector field described by ``most probable'' directions. This approach transforms the information from HOG into a more comprehensive representation of the image's structure.

If the dominant orientation across various windows in the image remains consistent or if the orientation changes smoothly in local regions, it is more likely that the candidate is TP, i.e. an actual defect. This is because consistent orientations suggest structural coherence, while irregular orientation changes could indicate noise or atypical regions, which are less likely to be defects. This is particularly relevant when defects are understood as cracks or breaks rather than bumps, where coherent orientation patterns are expected.

A Laplacian filter is applied to this matrix by convolving it with a $3 \times 3$ kernel to further emphasize abrupt orientation changes. The Laplacian operator helps detect areas of high curvature or sharp changes in orientation, making irregularities more detectable:
\[
\Delta I = I * L =  I *
\begin{bmatrix}
0 & 1 & 0 \\
1 & -4 & 1 \\
0 & 1 & 0
\end{bmatrix}.
\]

Then, the variance of the orientations (\(\sigma_{\theta}^2\)) is calculated to measure the degree of fluctuation or inconsistency of the orientation in the candidate image. Low variance suggests smooth and consistent orientation changes, which is more indicative of a typical defect (such as a crack or a break), while high variance implies abrupt and irregular changes in orientation, indicating noise or a potentially atypical region, making it less likely to be a defect.

Following this, the DFT is applied to the complex image formed by the HOG vectors. This transforms the vector-valued image from the spatial domain into the frequency domain, providing a frequency representation that captures periodic patterns and structures. The DFT is essential for analyzing the frequency characteristics of the image, which can be particularly useful for detecting defective parts. By examining the frequency components, it becomes possible to identify repeating patterns, anomalies, or irregularities that might indicate the presence of a defect.

Let \( I_{\text{HOG}}: \Omega \subseteq \mathbb{N^*}^2 \to \mathbb{R}^2 \) be the mapping that associates each pixel position \((x,y)\) in the image with a HOG vector \( I_{\text{HOG}}(x,y) \) in \( \mathbb{R}^2 \), where \( \Omega \) represents the set of pixel positions in the image. The DFT of the HOG vector-valued image is then computed as:
\[
\tilde{I}_{\text{HOG}}(u,v) = \sum_{x=1}^{M} \sum_{y=1}^{N} I_{\text{HOG}}(x,y) e^{-j2\pi \left( \frac{ux}{M} + \frac{vy}{N} \right)},
\]
where \( I_{\text{HOG}}(x,y) \) represents the value of the HOG vector at position \((x, y)\), and \( \tilde{I}_{\text{HOG}}(u,v) \) is the frequency representation of the image at spatial frequencies \( u \) and \( v \). After obtaining the DFT, we proceed to calculate the divergence and curl of the vector field using partial and cross-derivatives, employing Sobel filters to compute the gradients. These operators are defined as follows:
\[
\operatorname{div}(F(x, y)) = \nabla \cdot F(x, y)= \left[ \frac{\partial}{\partial x} \frac {\partial}{\partial y} \right] \cdot \begin{bmatrix} F_x \\ F_y \end{bmatrix}
= \frac{\partial F_x}{\partial x} + \frac{\partial F_y}{\partial y},
\]
\[
\operatorname{curl}(F(x, y)) = \nabla \times F(x, y) \left( \equiv \left| \begin{matrix} 
\frac{\partial}{\partial x} & \frac{\partial}{\partial y} \\
F_x & F_y 
\end{matrix} \right|
\right) = \frac{\partial F_y}{\partial x} - \frac{\partial F_x}{\partial y}.
\]

The divergence and curl provide significant insight into the structural characteristics of the candidate image. A highly positive or negative divergence indicates that the region is either contracting or expanding. If the divergence is significantly positive, it suggests that the region is predominantly pointing inward, while a significantly negative divergence indicates outward orientation; both cases could signify a potential defect in the candidate image. Additionally, a large curl value suggests abrupt or dynamic changes in orientation, which may indicate the presence of strong edges or contours. If the curl is high, it further supports the likelihood that the candidate is defective.

To extract features from the divergence and curl of the HOG descriptor field, two steps are used: calculating the $L_1$ and $L_2$ norms in each 2D vector field matrix:
\[
\|F\|_1 = \sum_{m,n} |F_{m,n}|, \quad
\|F\|_2 = \left( \sum_{m,n} |F_{m,n}|^2 \right)^{\frac{1}{2}}.
\]

These norms essentially measure the total variation in the vector field and the total energy in terms of Euclidean length, respectively.

From here, five characteristics emerge.

\subsubsection{LBP} 
\hspace*{0.25cm} A
The Local Binary Patterns (LBP) descriptor is used to identify distinctive patterns associated with defects by capturing the texture of the candidate region while remaining invariant to changes in local illumination. To standardize defect analysis, the candidate's histogram is enhanced using the Contrast Limited Adaptive Histogram Equalization (CLAHE) method, which improves contrast in images with irregular lighting without oversaturating extreme areas, unlike traditional histogram equalization. For each pixel:
\[
LBP(x,y) = \sum_{n=0}^{N-1} \mathbf{1}_{(I(x_n, y_n) - I(x,y) \geq 0)} \cdot 2^n,
\]
where \( I(x_n, y_n) \) is the intensity of neighboring pixels around \((x,y)\) at a distance \(r\), \(N\) is the number of neighboring pixels considered (e.g. 8 for a \(3 \times 3\) neighborhood), and \(\mathbf{1}_{(.)}\) returns 1 if the neighboring pixel's intensity is greater than or less than the center pixel's intensity.

From the resulting LBP matrix, its entropy is extracted using its histogram, as it provides a quantitative feature that represents the texture characteristics of the candidate region.

From here, one characteristic emerges.
\subsubsection{Matrix texture analysis} 
\hspace*{0.25cm} We finalize using various matrix texture analysis methods to capture spatial relationships and pixel intensity patterns in a candidate, providing a description of the texture based on amplitude transitions. Among existing techniques in the literature, some useful ones are:

\begin{itemize}[label={-}]
    \item {GLCM (Gray-Level Co-occurrence Matrix)}: It captures the spatial relationship between pairs of pixels by counting how often pairs of pixel intensity values (i.e., gray levels) occur in a given spatial relationship (distance and direction). The matrix is computed for several orientations, typically 0, 45, 90, and 135°, and it represents the number of times intensity $i$ co-occurs with intensity $j$ in a certain direction:
\[
\begin{aligned}
\text{GLCM}(i, j, d) &= \\
\sum_{x, y} \, &
\begin{cases} 
1, & \text{if } I(x, y) = i \quad \land \\
& I(x+d_x, y+d_y) = j \\
0, & \text{otherwise}.
\end{cases}
\end{aligned}
\]
    \item {OGCM (Orthogonal Gray-Level Co-occurrence Matrix)}: It is a variation of GLCM, but it focuses on orthogonal directions such as North-South (vertical) and East-West (horizontal):
\[
\begin{aligned}
\text{OGCM}(i, j) &= \\
\sum_{x, y} \, &
\begin{cases} 
1, & \text{if } I(x, y) = i \quad \land \\
& I(x+d_x, y+d_y) = j \\
0, & \text{otherwise}.
\end{cases}
\end{aligned}
\]
    \item {GLRM (Gray-Level Radius Co-occurrence Matrix)}: It extends GLCM by considering the radial distribution of gray levels around a central point in concentric rings. This matrix is calculated by examining co-occurrences of pixel intensities within defined radial distances from a central pixel: where $(x', y')$ are pixels within the radius $r$ of the central pixel:
\[
\begin{aligned}
\text{GLRM}(i, j, r) &= \\
\sum_{x', y'} \, &
\begin{cases} 
1, & \text{if } I(x', y') = i \quad \land \\
& I(x'+\Delta x, y'+\Delta y) = j \\
& \text{ with } \sqrt{\Delta x^2 + \Delta y^2} = r \\
0, & \text{otherwise}.
\end{cases}
\end{aligned}
\]
    \item {FLCM (Filtered Level Co-occurrence Matrix)}: It analyzes pixel intensity co-occurrences after applying various filters to an image. By modifying pixel values through filters that enhance specific features, such as edges, the FLCM captures texture information that reflects these alterations. It counts how often pairs of pixel intensities appear in defined spatial relationships within the filtered candidate I':
\[
\begin{aligned}
\text{FLCM}(i, j) &= \\
\sum_{x, y} \, &
\begin{cases} 
1, & \text{if } I'(x, y) = i \quad \land \\
& I'(x+d_x, y+d_y) = j \\
0, & \text{otherwise}.
\end{cases}
\end{aligned}
\]
    \item {HOG-C (Histogram of Oriented Gradients – Complex)}: As in a previous feature developed, this bidimensional matrix uses the HOG descriptor and maps gradient orientations of the candidate region into a complex plane, treating each pixel's gradient vector as a complex number where the magnitude represents gradient strength and the angle represents the gradient direction:
\[
\begin{aligned}
\text{HOG-C}(r, \theta) &= \\
\sum_{x, y} \, &
\delta\left( r - |G(x, y)| \right) \cdot \delta\left( \theta - \arg(G(x, y)) \right).
\end{aligned}
\]
    \item {HGCM (Histogram of Gradient Co-occurrence Matrix)}: It combines both gradient information and spatial co-occurrence relationships. It calculates co-occurrences between gradient magnitudes and directions in a manner similar to the GLCM but using gradient vectors:
\[
\begin{aligned}
\text{HGCM}(i, j) &= \\
\sum_{x, y} \, &
\begin{cases} 
1, & \text{if } |G(x, y)| = i \quad \land \\
& |G(x+d_x, y+d_y)| = j \\
0, & \text{otherwise}.
\end{cases}
\end{aligned}
\]
\end{itemize}

After computing all these matrices, multiple metrics are derived from them:
\[
\text{Correlation} = \frac{\sum_{i,j} (i - \mu_x)(j - \mu_y) p(i,j)}{\sigma_x \sigma_y},
\]
\[
\text{Energy} = \sum_{i,j} p(i,j)^2,
\]
\[
\text{Entropy} = - \sum_{i,j} p(i,j) \log(p(i,j)),
\]
\[
\text{Homogeneity} = \sum_{i,j} \frac{p(i,j)}{1 + |i - j|},
\]
\[
\text{Variance} = \sum_{i,j} (i - \mu)^2 p(i,j),
\]
where \(p(i,j)\) is the probability associated with the cooccurrence of intensities \(i\) and \(j\) in any of the proposed texture matrices.

From here, 30 characteristics emerge, 5 for each matrix.

\section{Statistical feature selection}
In this section, we will begin with an introductory motivation and then explain the proposed statistical framework. This framework involves selecting features that, in a Fisher-optimal sense, are the best among the overall set defined earlier.

\subsection{Motivation}
With many of the features presented, one might think of computing a simple classification scheme based on thresholding, comparing the median values of ratios between. One way to manually improve threshold-based classification accuracy is to dynamically manually adjust the threshold value to maximize the separation between the mean and median of the TP and FP distributions:
\[
A_{\min}^\star = \arg \max_{A_{\min}} \left\{ \left| \mu_{\text{TP}} - \mu_{\text{FP}} \right| + \left| \tilde{R}_{\text{TP}} - \tilde{R}_{\text{FP}} \right| \right\},
\]
where \( \mu_{\text{TP}} \), \( \mu_{\text{FP}} \) are the means of the isolated pixel ratios of TP and FP, and \( \tilde{R}_{\text{TP}} \), \( \tilde{R}_{\text{FP}} \) are the medians. This idea is the foundation of our work, motivating the development of a more sophisticated framework to extend and refine this approach for heuristically enhancing robustness and versatility of classification.

In this second stage, a single analysis flow is executed to retain the ``best features'' from all those previously extracted. The features selected in this process are the only ones that, based on the provided observations, are (in some sense) worthwhile distinguishing between TP and FP classes.

Each basic feature previously considered will be analyzed, evaluating the statistical distribution they tend to follow to ensure long-term reliability. The best features will be filtered on the basis of various criteria, in order to identify those that best separate defects. 

The main goal is to retain the features with the maximally separated means for FP and TP cases, enabling easy differentiation of defects using a simple threshold technique. Additionally, these features must meet certain conditions regarding their distribution and variability to ensure their effectiveness over the long term and across different scales. 

Selecting features with the most separated means or distributions between classes is essential for robust dichotomous classification in highly variable industrial settings. Defects can arise from many different causes, making their detection complex. By choosing features that maximize separation, we enhance the model's ability to distinguish defects from noise, reduce errors, and ensure the system remains adaptable to diverse defect types. This approach simplifies classification, making it more accurate and reliable, even when the characteristics of defects vary widely.

\subsection{Methodology}
This step involves a black-box-type process that picks the best features (outputs) from all the extracted ones (inputs). These selected features are the most useful for distinguishing between classes. 

\begin{figure}[h!]
    \centering
    \includegraphics[width=\linewidth]{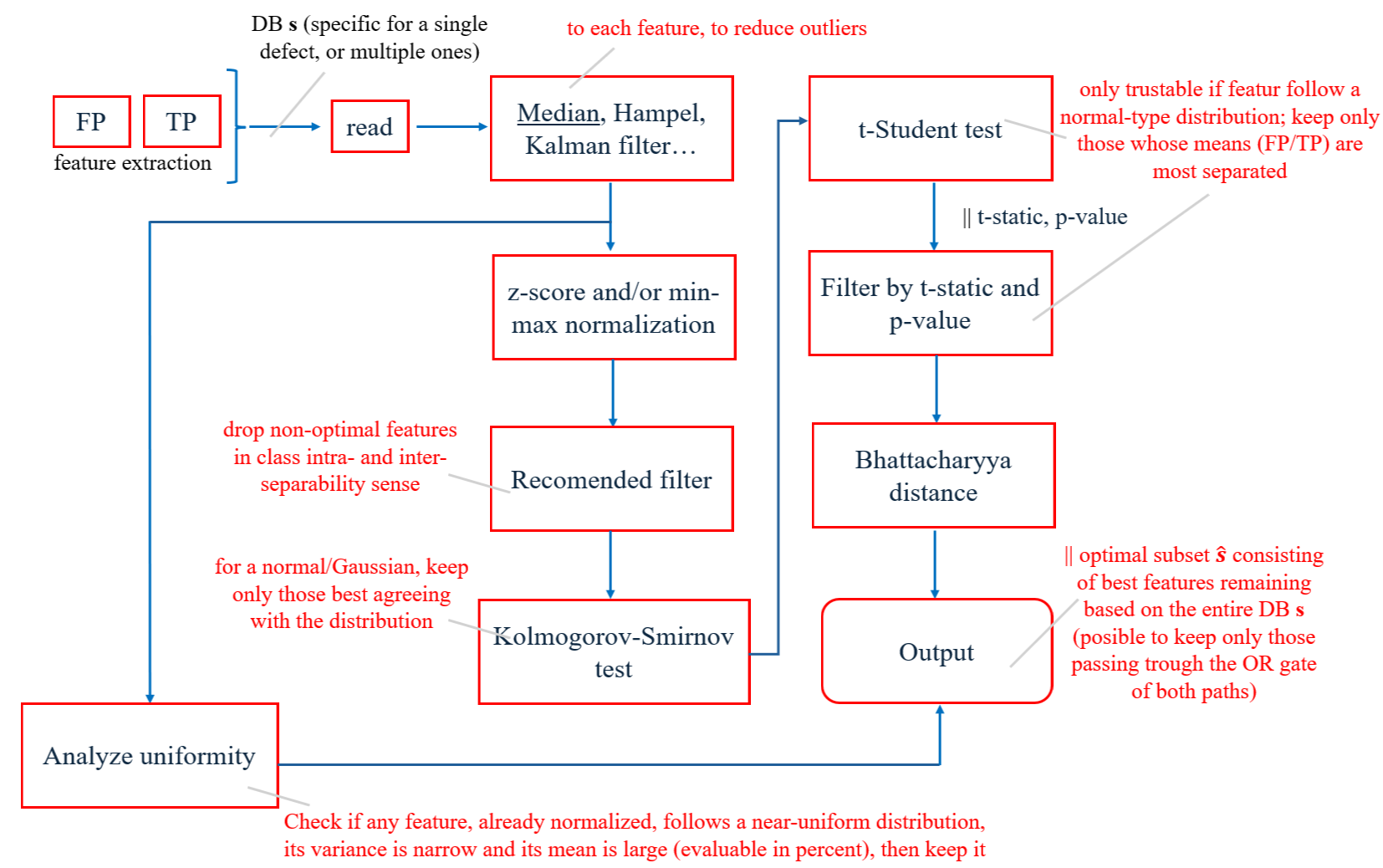}
    \caption{Basic workflow of the statistical selection stage.}
    \label{fig:methodology}
\end{figure}

The goal is to find which entry features separate TP from FP most effectively. The selected features should have well-separated means between the two classes and meet statistical conditions for long-term reliability. A visual intuition of the proposed method is shown in Fig.~\ref{fig:methodology}.

Two selection approaches are established: On the one hand, we look for features with a roughly uniform distribution, so features are normalized in a common frame, and only those with the most separated empirical means and sufficiently narrow variance are kept (approach A). Then, it is presumed that features should follow a normal distribution, so we keep only those maximizing between-class variability (separating TP/FP) while minimizing within-class variability (making features more distinct); we propose a concatenation of the K-S test, a Student's t-test, and the Bhattacharyya distance (B-distance) for this purpose (approach B).

In the following, a detailed explanation of the steps for both approaches is given. 

We start with two persistent matrices that contain the values of the feature set for each observation, \( M \) for TP and \( N \) for FP, \( X_{TP} \in \mathbb{R}^{N \times d} \) and \( X_{FP} \in \mathbb{R}^{M \times d} \). Each row of these matrices, \( x_i^{TP} \) or \( x_i^{FP} \), represents a feature vector for a given observation. For robustness, outliers are filtered using a Hampel filter of window size 5 \cite{Pearson2016}. 

For approach A, each feature is normalized based on its dynamic range and then the \( n \) features with the most separated empirical means between TP and FP and the narrow empirical variance are kept:
\[
x_i^{\text{norm}} = \frac{x_i - \min(x_i)}{\max(x_i) - \min(x_i)}.
\]

\begin{figure}[h!]
    \centering
    \includegraphics[width=\linewidth]{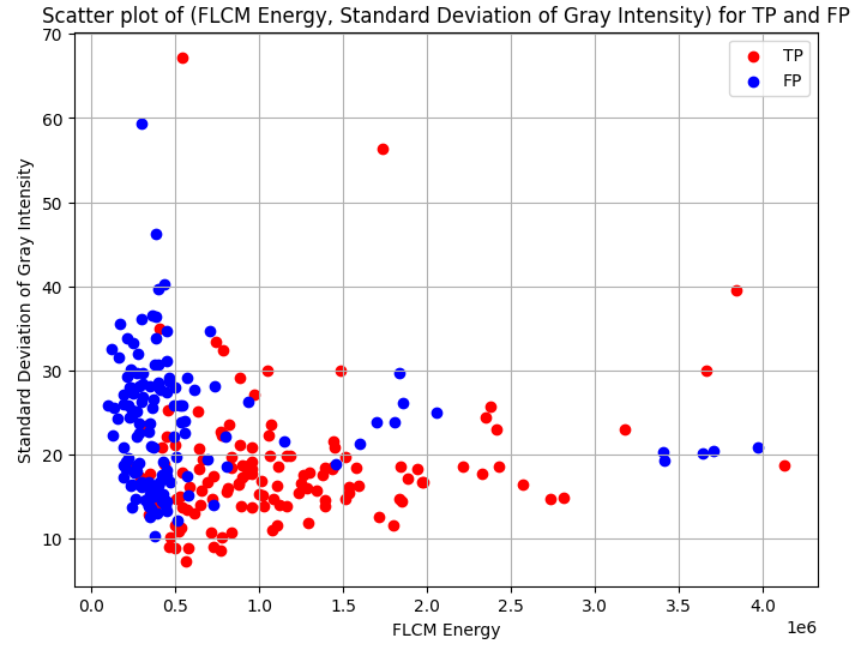}
    \caption{Graphical representation of two separable extracted features from stage A.}
    \label{fig:twofeatures}
\end{figure}

\begin{figure}[h!]
    \centering
    \includegraphics[width=\linewidth]{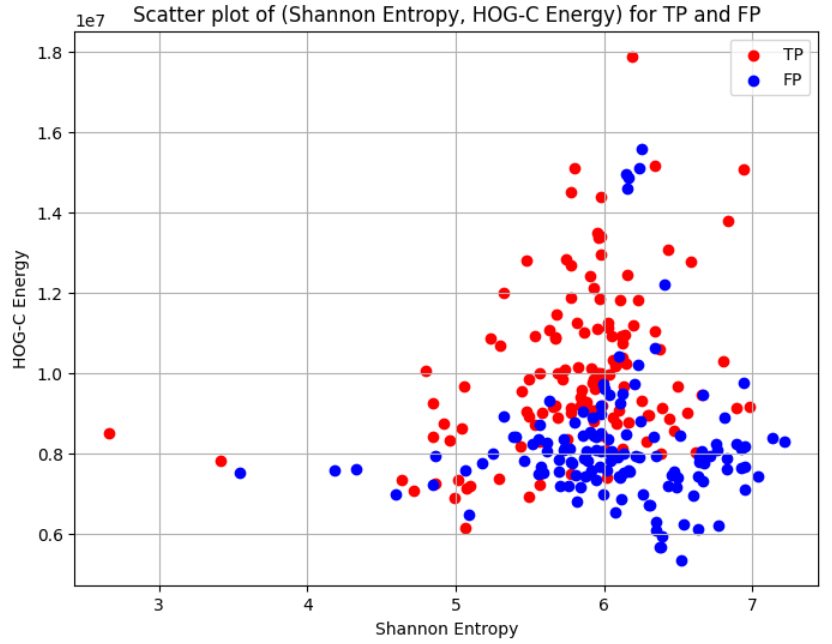}
    \caption{Graphical representation of two weakly separable extracted features from stage A.}
    \label{fig:twofeatures2}
\end{figure}

The idea of this normalization step is to normalize each feature by its absolute minimum and maximum values across both classes, ensuring comparable ranges. Intuitively, the characteristics that remain typically constant for each class are desired, that is, those that resemble a random variable following a uniform distribution \( X \sim U(a, b) \). The features whose variance is above a threshold (e.g., 0.8), once normalized in this way, are discarded; for a uniform distribution, if the following is true
\[ \frac{(b-a)^2}{12} > \tau, \quad \tau \in [0, 1], \]
then the feature is preserved and discarded otherwise. See Fig.~\ref{fig:twofeatures} and Fig.~\ref{fig:twofeatures2} for an example of two features that do and do not present clear inter-class linear separability.

For approach B, the selected features are then normalized, utilizing z-score normalization to maintain the relative scale between the samples while remaining transparent to the actual values' scale. 

Let \( \sigma_i \) denote the standard deviation of feature \( x_i \) and \( \text{cov}(x_i, x_j) \) denote the covariance between features \( x_i \) and \( x_j \). A very recommended feature \cite{Mery2015} is to remove features that have low variability (constant features), i.e., \( X' = X \setminus \{x_i\} \) if \( \sigma_i < \epsilon_1 \) (e.g., \( \epsilon_1 = 10^{-8} \)), where \( X' \) is the new feature matrix after suppressing all invalid \(x_i\). Also, it is suggested to eliminate one of the features from pairs that are highly correlated, i.e.:
\[
X' = X \setminus \{x_i\} \quad \lor \quad X' = X \setminus \{x_j\} \quad 
\]
\[
\text{if} \quad \left| \frac{\text{cov}(x_i, x_j)}{\sigma_i \sigma_j} \right| > \epsilon_2 \quad \text{(e.g., \( \epsilon_2 = 0.99 \))}.
\]

Subsequently, the K-S test is performed for a Gaussian/normal distribution \( \mathcal{N}(0, 1) \) to retain only those features (already normalized) that best fit this distribution. This test measures the maximum discrepancy between the empirical and theoretical distribution of the data; the null hypothesis states that the data follow the theoretical reference distribution, i.e., \( H_0: X \sim \mathcal{N}(\mu, \sigma^2) \). This means that only those features that meet this criterion for both classes are retained. This test is asymptotically valid when the sample space tends to be infinitely large, so the more samples available, the better.

The empirical cumulative distribution (CDF) of the empirical data and a normal are calculated:
\[
Empiric \text{ CDF} \equiv F_e(x) = \frac{N_{\text{data} \leq x}}{N_{\text{sample}}},
\]
where \(N_{\text{data} \leq x}\) is the number of data points less than or equal to \(x\) and \(N_{\text{sample}}\) is the total sample size. 

The CDF of a normal distribution is:
\[
Normal \text{ CDF} \equiv F_n(x) = \frac{1}{2} \, \text{erf}\left( -\frac{\mu_{\text{sample}}}{\sigma_{\text{sample}} \sqrt{2}} \right),
\]
where \(\mu_{\text{sample}}\) and \(\sigma_{\text{sample}}\) are the mean and standard deviation of the sample, respectively, and \(\text{erf}(z)\) is the error function, which provides the probability that a normally distributed random variable falls within the range of \([-z, z]\).

Following this, the K-S distance \(d_{K-S}\) is calculated, which measures the maximum difference between the empirical and theoretical CDFs:
\[
d_{K-S} = \max\left( \left| F_e(x) - F_n(x) \right| \right).
\]

Finally, for the K-S test, the p-value is calculated based on \(d_{K-S}\), which tractably quantifies the significance of the test:
\[
\text{p-value} \approx \text{erf}\left(d_{K-S} \sqrt{n}\right),
\]
where \(n\) is the sample size.

Then, using the features that are deemed ``more normal'', a Student's t-test is conducted, which is more reliable when the data follows a normal distribution. The test provides an additional statistical filter for the features, retaining only those for which the t-statistic exceeds a specified threshold or whose p-value falls below a certain level, thus rejecting the null hypothesis that posits no difference (separation) between the means of the two classes, hance ensuring that only features with sufficiently separated distributions are retained.

The t-statistic is calculated as:
\[
t = \frac{| \bar{x}_1 - \bar{x}_2 |}{\sqrt{\frac{\sigma_1^2}{n_1} + \frac{\sigma_2^2}{n_2}}}
\]
and the corresponding p-value can be approximated as \cite{Winitzki2008}:
\[
p - value \approx \frac{1}{2} \left( 1 - \text{erf} \left( \frac{t}{\sqrt{2}} \right) \right) \quad \text{(two-tailed test)}.
\]

It is important to note that the degrees of freedom (DoF) represent the number of independent values in a sample that can vary freely, subject to certain restrictions imposed by the data structure or hypothesis being tested, and are typically estimated using the Welch-Satterthwaite equation \cite{Shieh2015}. If the number of samples in both classes is equal, the degrees of freedom can be calculated as:
\[
\text{DoF} = n_1 + n_2 - 2.
\]
In general, a higher DoF leads to a t-distribution that approximates the standard normal distribution more closely.

Finally, for this approach, the B-distance between the classes is calculated on the basis of the set of the best features retained so far, which evaluates the classification quality for a given feature. The features are organized into two matrices \(X'\), \(Y'\) where each matrix contains as many rows as there are samples/observations and as many columns as there are features/variables. From these matrices, the mean vector \( \mu \) and the covariance matrix \( \Sigma \) are computed, which are then used to calculate the B-distance.

The equation for the Bhattacharyya distance between two classes can be expressed as:
\[
D_B = \frac{1}{8} \left( \mu_1 - \mu_2 \right)^T \left( \Sigma_1 + \Sigma_2 \right)^{-1} \left( \mu_1 - \mu_2 \right) \, +
\]
\[
\quad \frac{1}{2} \ln \left( \frac{\det(\Sigma_1 + \Sigma_2)}{\sqrt{\det(\Sigma_1) \det(\Sigma_2)}} \right).
\]

This equation is applicable when the features approximately follow a (multivariate) normal distribution \cite{Kashyap2016}, which is expected at this stage if any features were retained. Only features with a B-distance less than a specified threshold \( \tau \) are retained, i.e. \( X' = X \setminus \{x_i\} \) if \( D_B < t \) (e.g. \( t = 100 \)).

Furthermore, since there may be no prior knowledge regarding the probability of FP and TP, it may be assumed that they are equiprobable (i.e., \( P(FP) = P(TP) = 0.5 \)). Therefore, the Chernoff bound, which provides an indication of the classification reliability of the metric and is tabulated, could be utilized for the selection of features.

\section{Discussion}
The proposed framework introduces a computational entity, or black-box module, designed to process the output of another model (e.g., an ML model). This module takes inputs such as detected candidate ROI images and enables their classification into relevant categories such as noise or defects. Although this study does not include extensive experimental validation, we outline how the framework could be tested to evaluate its effectiveness and provide meaningful comparisons to existing models.

The framework has been applied in practice, where the parameters (weights) were heuristically tuned to address specific data typologies of defects. This ad hoc approach allowed domain experts to adjust the weights based on their knowledge of the dataset and the noise distribution. However, a systematic experimental setup could be conducted to validate the framework by assessing its ability to match or improve ground-truth results produced by a state-of-the-art machine learning model, such as YOLOv8. The goal of such validation would be to evaluate whether the framework can mitigate overdetection from the prior model while maintaining or surpassing its performance as measured by a confusion matrix.

For each analytical metric \( M_i \) discussed earlier, a scalar score \( S_i \!\in\! [0,1] \!\subset\! \mathbb{R} \) is generated, which represents the conditional probability of a ROI belonging to the noise or defect class, given the metric \( M_i \):
\[
S_i = P(\text{Class} \!\mid\! \text{ROI}, M_i),
\]
where Class \( \in \{ \text{Noise}, \text{Defect} \} \). Using Bayes' theorem, we can rewrite this probability as:
\[
S_i = \frac{P(M_i \!\mid\! \text{Class})P(\text{Class})}{P(M_i)}.
\]

This allows each score \(S_i\) to be derived from the likelihood \(P(M_i \!\mid\! \text{Class})\) and the prior probabilities \(P(\text{Class})\).

Next, these scores can be combined into a composite score \( S \) through a weighted summation:
\[
S = \sum_{i=1}^{n} w_i S_i,
\]
where \( w_i \) represents the weight assigned to each metric \(M_i\), determined by its relative importance in the classification process. The composite score \(S\) can be interpreted as the total probability of the class belonging to the ROI, thereby facilitating informed decision making in classification.

This simple initial approach allows for the ad hoc selection of scores, where analysts and experts in the field can manually adjust the weights \( w_i \) based on their experience and domain-specific knowledge. However, this process may be limited by the subjectivity and inherent variability in weight selection, which can affect the consistency and precision of the classification model.

To address the limitation of the proposed simple score-based approach, a random forests-based approach can be utilized, which is a robust and effective supervised learning method. By integrating random forests, we can automate the assignment of weights to each metric through model training. In this case, the model learns to empirically weigh analytical metrics \( S_i \) based on their contribution to the overall classification performance.

The structure of an RF is based on creating multiple decision trees, each trained on a random sample of the dataset. During this process, each tree evaluates the metrics \( S_i \) and determines which are the most informative for making accurate predictions. The result is an important assignment for each metric, represented by the frequency with which it is used in tree splits and its impact on reducing the impurity of the node.

The final output is a combination of the decisions of each tree, allowing the random forest model to generate a composite score \( S \) expressed as:
\[
S = \frac{1}{T} \sum_{t=1}^{T} h_t(\text{ROI}),
\]
where \( h_t = \{ 0, 1 \} \) is the prediction of the \( t \)-th tree and \( T \in \mathbb{N}^* \) is the total number of trees in the forest. This integration not only improves model precision, but also provides a more solid basis for interpreting the metric scores, eliminating reliance on ad hoc choices and instead relying on empirical evidence.
This technique enables us to learn the optimal weights for each analytical metric, thus improving classification accuracy.

We have presented a novel approach that leverages a wide range of analytical metrics, combining them into a simple yet ad hoc weighted scoring system, but one could further enhance and generalize this system to optimize decision making. Using these comprehensive features, learning methods, such as an RF classifier, could be employed, which enables the usage of optimal weights for each analytical metric, enhancing overall performance and mitigating overfitting. By averaging the results of multiple decision trees trained on random subsets of the data. Specifically, in an RF each tree votes on the class label, and the most frequent label is selected as the final prediction, which enhances model generalization. By recursively splitting the data based on decision rules that minimize impurity, an RF aggregates predictions through majority voting for classification tasks. 

The ideal methodology would include training the model on a subset of the image data and validating its performance on unseen data, ensuring that the model generalizes well beyond the training set. Thus, enabling the evaluation of the classification accuracy of the model using standard performance metrics, with the ultimate goal of achieving a high-performing, reliable image classifier for dichotomic image classification tasks. Future work may involve this type of experimentation and present results for each analytical metric, highlighting their individual performance in distinguishing noise from defects using metrics such as accuracy, F1 score, and ROC-AUC to evaluate the hybrid weighted approach.

This framework not only enhances and generalizes the functionality of a precisely tailored model for a specific dataset, but also enables the exploration of the entire feature set to identify even a single feature that performs exceptionally well. This feature could then be used with a simple thresholding method for practical implementation. In essence, the framework paves the way for a wide range of possibilities to study the classification problem with significant flexibility.

\section{Conclusions}
This work presents a hybrid approach that successfully combines multiple analytical, image processing, and statistical methods to improve the classification of noise in images. The integration of specifically designed manual steps with tailored features and statistical-based insights offers a robust approach for particular industrial inspection applications and complex pattern recognition tasks in noisy environments.

The study aims to improve the effectiveness of statistical feature selection methods and the precision of defect detection. By focusing on various specific properties of the image data, this method helps reducing FPs and ensures that only the most relevant features are used in any classifier, where These selected features can be seamlessly integrated, significantly enhancing their performance without requiring any retraining or parameter tuning for different defect types or production environments.

Given the nature of the images and defects analyzed, we have focused on luminosity, intensity distribution, and statistical pixel metrics, rather than contour extraction or shape matching. Although we present a subset of potential features, many others can be implemented at varying detailed levels. Transforming candidates into domains defined by basis functions, such as wavelets, enables multiscale analysis, thus enhancing robustness against noise.

In future research, we also aim to explore the application of Multilayer Perceptron (MLP) and boosting algorithms using the complete set of both analytical features. Leveraging the capacity of MLPs to capture complex relationships in the data, we might expect to enhance classification accuracy further. In addition, employing boosting techniques may improve model performance by sequentially addressing the weaknesses of initial predictions. This exploration would allow us to compare the effectiveness of these methods against an RF approach and assess how well they handle the full spectrum of features.

Other future efforts could focus on enhancing model performance through feature importance analysis, using an RF classifier to rank features according to their contribution to predictions. By training the model on all features, evaluating importance scores and excluding less relevant features, the model can be retrained using only the most impactful ones. This approach reduces the feature space, improving efficiency and potentially boosting accuracy. A comparison of the performance of the retrained model against the original would validate the benefits of this optimization.

It would be valuable to design a specific methodology for some or all metrics to guide the automatic parameterization of variables. This would eliminate the need for manual fine-tuning based on the dataset type or specific noise distribution, making the process less labor intensive and time consuming.
 
Moreover, it would be worthwhile to investigate advanced deep learning architectures and multiscale analysis techniques, including the Discrete Wavelet Transform (DWT) and Hadamard-Walsh transform. These methods could improve classification performance by enabling more robust and sophisticated feature extraction. In addition, more mature features can be identified and exploited in an aggregation-leaning approach, taking advantage of a large ``noisy'' dataset, even in a unsupervised manner.

\bibliography{export.bib}

\end{document}